\documentclass[12pt]{article}
\usepackage{amsmath}
\usepackage{graphicx}
\usepackage{enumerate}
\usepackage{natbib}
\usepackage{url} 
\usepackage{color}
\usepackage{amsfonts,exscale,relsize}
\usepackage{amssymb}
\usepackage[mathscr]{eucal}
\usepackage{amsthm,amsmath,bm} 
\usepackage{enumitem}
\usepackage{dsfont, multirow}
\usepackage{relsize}
\usepackage{appendix}

\graphicspath{{figures/}}

\newcommand{\blind}{1}

\newtheorem{assumption}{Assumption}
\newtheorem{remark}{Remark}[section]
\newtheorem{theorem}{Theorem}[section]
\newtheorem{prop}{Proposition}[section]
\newtheorem{lem}{Lemma}

\begin{document}
\begin{sloppypar}

\def\spacingset#1{\renewcommand{\baselinestretch}%
{#1}\small\normalsize} \spacingset{1}

\def\MLE{\mbox{\smallfont{MLE}}}
\def\U{\mbox{\smallfont{U}}}
\def\P{\mbox{Pr}}


\if1\blind
{
  \title{\bf A unified framework for bounding causal effects on the always-survivor and other populations}
  \author{Aixian Chen\\
    Department of Economics and Statistics, Guangzhou University, China\\
    and \\
    Xia Cui \thanks{
    \textit{This work was supported by the National Natural Science Foundation of China (Grant No. 11871173)}}\hspace{.2cm}\\
    Department of Economics and Statistics, Guangzhou University, China\\
    and\\
    Guangren Yang\\
    Department of Economics, Jinan University, China}
    \date{}
  \maketitle
} \fi

\if0\blind
{
  \bigskip
  \bigskip
  \bigskip
  \begin{center}
    {\LARGE\bf A unified framework for bounding causal effects on the always-survivor and other populations}
\end{center}
  \medskip
} \fi

\bigskip
\begin{abstract}
We investigate the bounding problem of causal effects in experimental studies in which the outcome is truncated by death, meaning that the subject dies before the outcome can be measured. Causal effects cannot be point identified without instruments and/or tight parametric assumptions but can be bounded under mild restrictions. Previous work on partial identification under the principal stratification framework has primarily focused on the `always-survivor' subpopulation. In this paper, we present a novel nonparametric unified framework to provide sharp bounds on causal effects on discrete and continuous square-integrable outcomes. These bounds are derived on the `always-survivor', `protected', and `harmed' subpopulations and on the entire population with/without assumptions of monotonicity and stochastic dominance. The main idea depends on rewriting the optimization problem in terms of the integrated tail probability expectation formula using a set of conditional probability distributions. The proposed procedure allows for settings with any type and number of covariates, and can be extended to incorporate average causal effects and complier average causal effects. Furthermore, we present several simulation studies conducted under various assumptions as well as the application of the proposed approach to a real dataset from the National Supported Work Demonstration.
\end{abstract}

\noindent%
\vfill

\newpage
\spacingset{1.9} 

\makeatletter
\@addtoreset{equation}{section}
\makeatother
\renewcommand{\theequation}{\arabic{section}.\arabic{equation}}
\section{Introduction}
\addtolength{\textheight}{.5in}

The problem of `truncation by death' (see, for example, \citealp{2002Frangakis}) , often arises when the subject has died before the outcome could be measured. This may lead to flaws in causal analysis since a direct comparison between the `treated group' and the `control group' among only the observed survivors will result in selection bias. For example, in the estimation of the effects of medical treatments in clinical trials, the health outcomes for participants who have died are undefined.

To cope with the bias resulting from `truncation by death', \citet{2002Frangakis} proposed principal stratification to define the `survivor average causal effect' (SACE), which is a treatment comparison in the subpopulation of subjects who would survive under both treatment and nontreatment. In the absence of strong untestable parametric restrictions or instruments, causal effects cannot be point identified since the principal strata of interest cannot be observed directly. However, upper and lower bounds can still be obtained under fairly mild restrictions. In the context of `truncation by death' and/or `noncompliance', most literature has focused on identifying the average causal effect (ACE) on a subpopulation or the entire population, and two assumptions have often been imposed, either separately or jointly, to derive bounds on the ACE; these assumptions are (i) monotonicity of selection in the treatment and (ii) stochastic dominance of the potential outcomes of the 
`always-survivor' subpopulation over those of other populations. \citet{2003Zhang} (see also \citealp{2008Zhang}) derived large sample bounds on the ACE for the `always-survivor' subpopulation, but their bounds involved numerical optimization in some observed subpopulations. \citet{2008Imai} used another method to prove that the bounds are sharp, and simplified them into closed-form expressions. \citet{2014Freiman} extended these bounds to the case with binary covariates under weakly ignorable treatment assignment. \citet{2009Lee} assessed the wage effect of the Job Corps program on the `always-survivor' subpopulation under monotonicity of selection and proved that the bounds are sharp. \citet{2011Blanco} considered the same program under mean dominance assumptions within and across subpopulations and obtained  tighter bounds. Assuming nonparametric/semiparametric models on outcome, \citet{2011Ding}
identified the SACE by differentiating conditional distributions between two principal strata via a substitution variable for the latent survival type. \citet{2014Tchetgen} identified the SACE by using the longitudinal correlation between survival and outcome after treatment under the monotonicity assumption. 
Following \citet{2011Ding}, \citet{2017Wang} relaxed the assumptions and bounded the SACE without covariates. For multivalued ordinal treatments, \citet{2021Luo} identified the SACE in two ways: by using some auxiliary variable and on the basis of a linear model. In presence of `truncation by death' (or sample selection) and noncompliance, \citet{2015Chen}, \citet{2019Kennedy} and \citet{2020Blanco} further considered the identification of the ACE for the survivor-complier subpopulation under instrument monotonicity and monotonicity of selection.


Another frequently adopted assumption is that the outcome is either discrete and finite or continuous and bounded. In their seminal work, \citet*{1997Balke} derived the tightest bounds over causal effects by employing an algebraic program to derive analytic expressions under discrete and finite outcomes. \citet{2012Beresteanu} showed that through the use of random set theory, the results of \citet[Corollary 2.2.1]{2003manski} and \citet*{1997Balke} can be simplified and extended. 
\citet{2015Shan} derived bounds on the SACE by applying the symbolic Balke–Pearl linear programming method under monotonicity when the outcome is binary. \citet{2017Huber} focused on partial identification of treatment effects on further subpopulations under noncompliance, in particular complier average causal effect (CACE) when the outcome is continuous and bounded. \citet{2020Gunsilius} bounded the ACE with continuous outcomes allowing for more than two-arm treatments and a continuous instrumental variable. In addition, \citet{2021Zhang} studied the partial identification of the ACE when the outcome is continuous and bounded in high-dimensional contexts. \citet{2021Kitagawa} provided closed-form expressions for the identified sets of the potential outcome distributions and ACE by allowing the outcomes to be continuous and bounded. \citet{2022Sachs} established two algorithms for deriving constraints and obtaining symbolic bounds that are valid and tight. However, the linear programming method of \citet*{1997Balke} cannot be used directly for outcome-dependent sampling designs. \citet{2022Gabriel} proposed two approaches to derive nonparametric bounds for the ACE with binary treatments and outcomes. The (partial) identification of the ACE was comprehensively reviewed by \citet{2018Swanson}.


However, in many practical applications, the monotonicity and stochastic dominance assumptions have not been empirically tested, and the meanings of monotonicity and dominance themselves are unclear. Furthermore, the outcome variable often has an infinite range and is therefore unbounded. For example, when assessing how training activity affects labor market success, such as employment or earnings across various subpopulations, the outcome is often assumed to be drawn from lognormal distribution. The goal in this paper is to investigate the partial identification of the ACE on discrete and continuous square-integrable outcomes that are truncated by death. More specifically, our contributions are as follows: (1) We present a unified way, allowing for either discrete or continuous outcomes, to calculate sharp bounds on the ACE using a linear Balke–Pearl program. The main idea depends on rewriting the optimization problem in terms of the integrated tail probability expectation formula using a set of conditional probability distributions. (2) We derive sharp bounds on the ACEs among the `always-survivor' and other subpopulations, namely, the `protected' and `harmed' subpopulations. (3) We compute the bounds on the ACE when some of the identifying assumptions are relaxed, for example, in the absence of monotonicity and stochastic dominance.These bounds allow for settings with covariates of any type. (4) We extend the framework to obtain the sharp bounds of ACE and CACE. Finally, our results are validated based on simulations and a real application on the National Supported Work Demonstration (NSW) data \citep{1986LaLonde,2020Sant}. 

The remainder of this article is organized as follows. Section 2 characterizes the `truncation by death' problem based on principal stratification. Section 3 discusses partial identification of the ACE for the always-survivor and observed (`protected' and `harmed') subpopulations under no assumptions, corresponding to the worst-case bounds. Section 4 considers bounding the SACE under the assumptions of monotonicity and/or stochastic dominance. Section 5 provides the estimation procedure of the resulting bounds.
In Section 6, we report several simulation studies conducted to evaluate the finite-sample performance of the proposed approach. In Section 7, we consider an empirical application to experimental data from NSW. We present conclusions in Section 8. The proposed procedure is extended to the case of ACE and CACE in \ref{Appendix: extension}.

\section{The `truncation by death' problem}
Suppose that we wish to bound the effect of a binary treatment, $Z_i\in\{0,1\}$, on an outcome $Y_i$ some time after assignment. Here, $Z_i=1$ indicates treatment and $Z_i=0$ indicates controlled. The principal stratification framework \citep{2002Frangakis} is used to motivate the `truncation by death' problem. Let $S_i(z)$ denote a subject’s potential survival outcomes under treatment $Z=z$ (1 for survival, 0 for death), i.e., $S_i(1)$ and $S_i(0)$ represent the survival status of a subject at follow-up under $Z=1$ and $Z=0$, respectively. 
We similarly denote the outcome of interest by $Y_i(z)$ under treatment $Z=z$, where $Y_i(1)$ and $Y_i(0)$ are the two potential outcomes that the subject would exhibit under treatment and nontreatment, respectively. Notably, the potential outcomes $Y_i(z)$ is defined only if $S_i(z)=1$; otherwise, they are undefined because of truncation by death. Let $W$ denote the covariate vector. The population can be divided into four principal strata, denoted by $G$, the definitions of which are summarized in Table~\ref{Table1}. This table shows that there is a one-to-one mapping relationship between the survival type $G$ and the bivariate latent survival status $\{S(1),S(0)\}$; therefore, $G$ can be understood as an abbreviation for $\{S(1),S(0)\}$. 
 
\begin{table}[htb]

 \renewcommand{\arraystretch}{0.8}
 \centering\caption{Subject survival types}
 \label{Table1}
 \scalebox{0.8}{
 \begin{tabular}{cccc|c}
           \hline
           $S_i(1)$  &  $S_i(0)$  & Survival type  & $G_i$ & Description \\
           \hline
1 & 1 & always-survivor&$g_0$&The subject always survives, regardless of the assigned treatment\\
1 & 0 & protected & $g_1$ & The subject survives if treated but dies if not treated (control)\\
0 & 1 & harmed & $g_2$ & The subject survives if not treated (control) but dies if treated\\
0 & 0 & doomed & $g_3$ & The subject always dies, regardless of the assigned treatment\\
           \hline
\end{tabular} }
\end{table}
The axiom of consistency \citep{2000Pearl} is adopted, such that the observed survival status $S$ and the observed outcome $Y$ satisfy  
\begin{equation*}
\begin{split}
 &S_i = Z_iS_i(1) + (1-Z_i)S_i(0),\\
 &Y_i = Z_iY_i(1) + (1-Z_i)Y_i(0),\,\,\mbox{if $S_i = 1$ and not observed otherwise.}
\end{split}
\end{equation*}
In practice, one of the two potential outcomes is observed if $S_i=1.$ A natural but crude temptation is to measure causal effects by comparing the means of $Y$ in each treatment arm among the observed survivors:
\begin{equation*}
E\{Y(1) - Y(0) \mid S = 1\}. 
\end{equation*}
Note that $S(1)=1$ involves two strata, $g_0$ and $g_1$, while $S(0)=1$ involves $g_0$ and $g_2$. Therefore, direct comparison between the `treated group' and the `control group' among those who survived ($S=1$) is, in fact, unfair. As an alternative, we define the ACE in a potential strata:
\begin{equation}\label{ace}
\Delta_k = E\{Y(1) - Y(0)  \mid G=g_k \}, \quad\mbox{ for $k=0,1,2,3$.}
\end{equation}
We do not consider bounds for the $G=g_3$ strata here because in this strata, the subject always dies regardless of the assigned treatment, which leads to no valid information from the observations. For $G=g_0$, $\Delta_{0}$ coincides with the SACE mentioned above. We first investigate the bounds of the SACE and then discuss the average treatment effects on other strata. The ACE (\ref{ace}) cannot be point identified without further assumptions since the principal strata of interest cannot be observed directly. However, upper and lower bounds can still be obtained under fairly mild restrictions. To place bounds on causal effects using the observed data, we assume that there is no interference between units, which means that the potential outcomes and potential survival statuses of one subject do not depend on the treatment statuses of other subjects and that there is only one version of treatment \citep{1980Rubin}.

\begin{assumption}[Stable Unit Treatment Value Assumption, SUTVA]\label{assu1} For individual $i$ and individual $j$, $Y_i(z) \perp Z_j$ and $S_i(z) \perp Z_j$, $\forall i\neq j, z\in \{0,1\}$, where ``$\perp$'' denotes independence.
\end{assumption}

If randomly assigned in causal inference, the treatment $Z$ is independent of the potential values of the survival status $S$ and outcome $Y$. When there are covariates, randomization is assumed to hold only conditional on the covariates $W$. We assume that the joint distribution of the potential outcomes and survival statuses is independent of the treatment given $W$ \citep{2000Pearl, 2015Huber, 2021Luo}:
\begin{assumption}[Ignorability on Observables]\label{assu2}
$Z_i \perp \{Y_i(1),Y_i(0),S_i(1),S_i(0)\} \mid W,G,$ with $G\in\{g_0,g_1,g_2,g_3\}$.
\end{assumption}

When there are covariates, we define $\pi_{k\cdot w} = \P(G = g_k \mid W=w)$ for $k=0,1,2,3$ and $P_{s \cdot zw} = \P(S=s \mid Z=z,W=w)$.  Under Assumption~\ref{assu2}, we can discover the relationship between the proportions $\pi_{k \cdot w}$ of the principal strata that are latent and the observed conditional probability $P_{s \cdot zw}$; the results are shown in Table~\ref{table_mix}. 
\begin{table}[htb]
 \renewcommand{\arraystretch}{0.8}
 \centering\caption{Observed probability and principal strata proportions}
 \label{table_mix}
 \begin{tabular}{cc}
           \hline
            Observed probability &  principal strata proportions \\
           \hline
$P_{1 \cdot 1w} = Pr(S=1 \mid Z=1,W=w)$ & $\pi_{0\cdot w}+\pi_{1\cdot w}$\\
$P_{1 \cdot 0w} = Pr(S=1 \mid Z=0,W=w)$ & $\pi_{0\cdot w}+\pi_{2\cdot w}$\\
$P_{0 \cdot 1w} = Pr(S=0 \mid Z=1,W=w)$ & $\pi_{2\cdot w}+\pi_{3\cdot w}$\\
$P_{0 \cdot 0w} = Pr(S=0 \mid Z=0,W=w)$ & $\pi_{1\cdot w}+\pi_{3\cdot w}$\\
           \hline
\end{tabular}
\end{table}

The following proposition whose proof is completed in Appendix \ref{Appendix: pi_k} provides the sharp bounds of quantities $\pi_{k \cdot w}$. 
\begin{prop} 
\label{prop1}
Under Assumptions \ref{assu1}-\ref{assu2}, the proportion of strata $g_2$ is bounded as 
\begin{equation}\label{b_pi0}
   \max\{0,P_{1\cdot 0w}-P_{1\cdot 1w}\}\leq \pi_{2 \cdot w} \leq \min\{P_{1\cdot 0w}, P_{0\cdot 1w}\}.
\end{equation}
Similarly, the following inequalities represent probability bounds in other principal strata:
\begin{eqnarray}
\begin{aligned}
  \max\{0,P_{1\cdot 1w}-P_{0\cdot 0w}\}\leq \pi_{0\cdot w} \leq \min\{P_{1\cdot 0w}, P_{1\cdot 1w}\},\\
  \max\{0,P_{1\cdot 1w}-P_{1\cdot 0w}\}\leq \pi_{1\cdot w} \leq \min\{P_{0\cdot 0w}, P_{1\cdot 1w}\}.
\end{aligned}
\end{eqnarray}
\end{prop}

The observed $(Z_i,S_i)$ generates the following two observed subgroups, which are mixtures of two principal strata, as presented in Table~\ref{sub}. 
\begin{table}[htb]
 \renewcommand{\arraystretch}{0.8}
 \centering\caption{Observed subgroups and principal strata}
 \label{sub}
 \begin{tabular}{c|c}
           \hline
           Observed subgroups  &  Principal strata  \\
           \hline
\{$i: Z_i=1, S_i=1$\} & Subject $i$ belongs either to $g_0$ or to $g_1$\\
\{$i: Z_i=0, S_i=1$\} & Subject $i$ belongs either to $g_0$ or to $g_2$\\
\hline
\end{tabular}
\end{table}

Unfortunately, even under Assumptions \ref{assu1} and \ref{assu2}, given $W$, point identification of either the principal strata proportion $\pi_{k \cdot w}$ or the conditional mean outcome within any strata, $\mu_{z,k,w}=E(Y \mid Z=z, G=g_k, W=w)$, is not possible. Instead, the observed conditional mean outcome is a mixture of the mean outcomes for two strata. For example,
\begin{align*}
    E(Y \mid Z=1, S=1, W=w) = \sum_{k=\{0,1\}}\frac{\pi_{k\cdot w}}{\pi_{0 \cdot w}+\pi_{1 \cdot w}}  \mu_{z,k,w}.    
\end{align*}
However, with the observed data, we can still obtain some useful information on bounds by invoking further assumptions such as monotonicity and stochastic dominance.

\section{Nonparametric bounds without further assumptions} 
\label{section:withoutM&SD}

Let $R$ be the potential responses of $Z$ on $Y$, which are divided into four types for $y\in R$ as follows:
\begin{eqnarray*}
\left\{
		\begin{aligned}
                r_0: \{Y(1)\leq y \cap Y(0) \leq y\}, \\
                r_1: \{Y(1)\leq y \cap Y(0) > y\}, \\
                r_2: \{Y(1) > y \cap Y(0) \leq y\}, \\
                r_3: \{Y(1) > y \cap Y(0) > y\} .
        \end{aligned}
\right.                
\end{eqnarray*}
Similarly, the potential responses of $Z$ on the survival status $S$ is divided into four types, consistent with the principal strata framework given in Table~\ref{Table1}. Moreover, we define the joint probability distribution of $R$ and $G$ conditional on $W$ as follows:            
\begin{equation} 
\label{define Fij|w }
F_{ij \cdot w} \equiv \P(R=r_i, G=g_j\mid W=w),\,\,\,  i,j=0, 1, 2, 3.
\end{equation}
The conditional SACE is defined as
\begin{align*}
\mbox{SACE}_w = \mu_{1,0,w}-\mu_{0,0,w}= E\{Y|Z=1,G=g_0,W=w\}-E\{Y|Z=0,G=g_0,W=w\}.
\end{align*}
Note that the conditional expectation of any integrable random variable $Y$ takes the form of an integral of its survival function, called the integrated tail probability expectation formula \citep{2018Ambrose}. Then, with $z=0,1,$
\begin{align*}
 &E\{Y|Z=z,G=g_0,W=w\}\\
 =&\int_{0}^{+\infty} \Big\{1-F(y|Z=z,G=g_0,W=w)\Big\}dy-\int_{-\infty}^{0} F(y|Z=z,G=g_0,W=w) dy.
\end{align*}
Thus,
\begin{eqnarray}\label{SACEw}
\begin{aligned}
\mbox{SACE}_w&= \int_{-\infty}^{+\infty}\Big \{F(y \mid Z=0, G=g_0, W=w) - F(y \mid Z=1, G=g_0, W=w) \Big\} dy,\\
&=\int_{-\infty}^{+\infty}  \frac {F_{20 \cdot w} -F_{10 \cdot w} }{F_{00 \cdot w} + F_{10 \cdot w} + F_{20 \cdot w} +F_{30 \cdot w}}  dy.
\end{aligned}
\end{eqnarray}
A natural idea arises that this is what we are solving for instead of deriving the bounds on $\mbox{SACE}_w$ directly. Hence, it is sufficient to bound the integrand term for any fixed $y\in R$.

Now, we further obtain some information about the conditional joint probability distribution $F_{ij\cdot w}$. Let $F_{as\cdot zw}$ denote the observed conditional joint probability distribution of $Y$ and $S=s$ given $Z=z$ and $W=w$, that is, 
\begin{equation}
F_{as\cdot zw}\equiv \P(I(Y\leq y)=a,S=s \mid Z=z,W=w),  a,s,z\in\{0,1\}.
\end{equation}
On the basis of the consistency property \citep{2000Pearl,2017Wang} and Assumptions \ref{assu1} and \ref{assu2}, the following constraints can be obtained:
\begin{eqnarray}
\label{linear programming 1}
\left\{
		\begin{aligned}
            &F_{11 \cdot 0w}=F_{00 \cdot w}+F_{20 \cdot w}+F_{1 \cdot 2w},\\
            &F_{01 \cdot 0w}=F_{10 \cdot w}+F_{30 \cdot w}+F_{2 \cdot 2w},\\
            &F_{11 \cdot 1w}=F_{00 \cdot w}+F_{10 \cdot w}+F_{0 \cdot 1w},\\
            &F_{01 \cdot 1w}=F_{20 \cdot w}+F_{30 \cdot w}+F_{3 \cdot 1w},\\
            &1=F_{00 \cdot w}+F_{10 \cdot w}+F_{20 \cdot w}+F_{30 \cdot w}+F_{0 \cdot 1w}+F_{3 \cdot 1w}+F_{1\cdot 2w}+F_{2\cdot 2w}+F_{\cdot 3w}, \\
            &F_{i0 \cdot w} \geq 0 \,(i=0,1,2,3), F_{0\cdot 1w}\geq 0, F_{3\cdot 1w}\geq 0, F_{1\cdot 2w}\geq 0, F_{2\cdot 2w}\geq 0, F_{\cdot 3w}\geq 0,
        \end{aligned}
\right.                
\end{eqnarray}
where $ F_{0\cdot 1w}=F_{01\cdot w}+F_{11\cdot w}, F_{3\cdot 1w}=F_{21\cdot w}+F_{31\cdot w}, F_{1 \cdot 2w}=F_{02\cdot w}+F_{22\cdot w}, F_{2 \cdot 2w}=F_{12\cdot w}+F_{32\cdot w}$ and $F_{\cdot 3w}=F_{03\cdot w}+F_{13\cdot w}+F_{23\cdot w}+F_{33\cdot w}$. The detailed derivation of (\ref{linear programming 1}) is given in Appendix \ref{Appendix: equ.(3.4)}, and the term $F_{a1\cdot zW}$ defined above is identifiable from the observed data. In order to obtain bounds of $\mbox{SACE}_w$ and SACE, it suffices to establish bounds of the following term subject to (\ref{linear programming 1}):
\begin{equation}\label{fracP}
    \frac {F_{20\cdot w}-F_{10\cdot w}}{F_{00 \cdot w} + F_{10 \cdot w} + F_{20 \cdot w}+F_{30 \cdot w}}.
\end{equation}
This optimization problem is not easy to directly solve since the objective function is nonlinear and the number of constraints in (\ref{linear programming 1}) is fewer than that in the situation without monotonicity and stochastic dominance \citep{2008Cai,2015Shan}. 

The sharp bounds of $\mbox{SACE}_w$ and SACE will be
established in the following theorem.
We need to introduce some notation. Let $\pi_{2\cdot w}^{\scriptstyle\max}$ denote the maximum value of $\pi_{2\cdot w}$ with $\pi_{2\cdot w}^{\scriptstyle\max}=\min\{P_{1\cdot 0w}, P_{0\cdot 1w}\}.$ Define
\begin{eqnarray*}
    \begin{aligned}
&R_0(y,w|\pi_{2\cdot w})= \min \left\{
\frac {F_{01 \cdot 1w}}{P_{1\cdot 0 w}-\pi_{2\cdot w}},
 \frac {F_{11 \cdot 0w}}{P_{1\cdot 0 w}-\pi_{2\cdot w}},
 \frac {F_{01 \cdot 1w}+F_{11 \cdot 0w}}{P_{1\cdot 0 w}-\pi_{2\cdot w}} - 1,
 1
\right\},\\
&L_0(y,w|\pi_{2\cdot w})= \max \left\{
     \frac {-F_{11 \cdot 1w}}{P_{1\cdot 0 w}-\pi_{2\cdot w}}, 
\frac {-F_{01 \cdot 0w}}{P_{1\cdot 0 w}-\pi_{2\cdot w}},
 \frac {-(F_{01 \cdot 0w}+F_{11 \cdot 1w})}{P_{1\cdot 0 w}-\pi_{2\cdot w}} + 1,
       -1
\right\},\\
        &L_0=\frac {E\Big\{(P_{1\cdot 0 W}-\pi_{2 \cdot W}^{\scriptstyle\max})\int_{-\infty}^{+\infty}L_0(y,W|\pi_{2\cdot W}^{\scriptstyle\max})dy\Big\}}{E(P_{1\cdot 0 W}-\pi_{2 \cdot W}^{\scriptstyle\max})},\\
        &R_0=\frac {E\Big\{(P_{1\cdot 0 W}-\pi_{2 \cdot W}^{\scriptstyle\max})\int_{-\infty}^{+\infty}R_0(y,W|\pi_{2\cdot W}^{\scriptstyle\max})dy\Big\}}{E(P_{1\cdot 0 W}-\pi_{2 \cdot W}^{\scriptstyle\max})}.
    \end{aligned}
\end{eqnarray*}
We can now state the result of sharp bounds on $\mbox{SACE}_w$ and SACE. 
\begin{theorem} \label{thm-SACE}Under Assumptions \ref{assu1}-\ref{assu2}, given covariates $W=w,$ the $\mbox{SACE}_w$ is bounded as
\begin{equation}\label{bound_00}
 \int_{-\infty}^{+\infty} L_0(y,w|\pi_{2\cdot w}^{\scriptstyle\max})\, dy\leq \mbox{SACE}_w \leq \int_{-\infty}^{+\infty} R_0(y,w|\pi_{2\cdot w}^{\scriptstyle\max})\, dy,
\end{equation} 
and the SACE is bounded as
\begin{equation}\label{bound_00_SACE}
 L_0\leq \mbox{SACE} \leq R_0.
\end{equation}
\end{theorem}
The proof of Theorem~\ref{thm-SACE} is relegated to Appendix \ref{Appendix: thm-SACE} based on the idea of the symbolic Balke-Pearl linear programming method. Like \citet{2014Freiman} that discussed large sample bounds for SACE with binary covariates, $\pi_{2\cdot w}$ is treated as a known parameter in the linear programming.
Theorem~\ref{thm-SACE} provides the bounds allowing for either discrete or continuous square-integrable outcomes, and any type of covariates. The resulting bounds are a covariate-adjusted version of those previously given by \citet{2003Zhang}. Covariates are often useful in narrowing the bounds, as illustrated by the simulation study in Section~\ref{sec:simu}.

Maximizing or minimizing (\ref{fracP}) subject to (\ref{linear programming 1}) leads to the bounds of (\ref{fracP}) are $[L_0(y,w|\pi_{2\cdot w}),R_0(y,w|\pi_{2\cdot w})].$ We will verify this from another perspective. 
Note that the domain of feasible solutions of (\ref{fracP}) is nonempty, that the denominator of (\ref{fracP}) does not reduce to a constant, and that it is strictly positive in the domain. According to Lemma~\ref{lemma: equivalence} in Appendix \ref{Appendix: lemma}, 
the fractional programming problem in (\ref{fracP}) has the same optimal solutions if and only if
$L_0(y,w|\pi_{2\cdot w}^{\scriptstyle\max})$ is the unique zero of the following parametric linear programming problem subject to (\ref{linear programming 1}),
\begin{equation*}
   \min \{F_{20\cdot w}-F_{10\cdot w}-\lambda (F_{00 \cdot w} + F_{10 \cdot w} + F_{20 \cdot w}+F_{30 \cdot w})\},
\end{equation*}
and $R_0(y,w|\pi_{2\cdot w}^{\scriptstyle\max})$ is the unique zero of 
\begin{equation*}
   \max \{F_{20\cdot w}-F_{10\cdot w}-\lambda (F_{00 \cdot w} + F_{10 \cdot w} + F_{20 \cdot w}+F_{30 \cdot w})\}.
\end{equation*}
\begin{remark}
We use the above algorithm to illustrate the bound $R_0(y,w|\pi_{2\cdot w}^{\scriptstyle\max})$ in detail. Let $\P(G=g_0\mid W=w)=\P(G=g_1\mid W=w)=0.4$ and 
$\P(G=g_2\mid W=w)=\P(G=g_3\mid W=w)=0.1$. We generate $Y$ from the following normal distribution conditional on $\{Z, G, W\}$: $Y \mid (Z=1, G=g_0, W=w) \sim N(7,1),\, Y \mid (Z=0, G=g_0, W=w) \sim  N(3,1),\,Y \mid (Z=1, G=g_1, W=w) \sim  N(5,1), \,Y \mid (Z=0, G=g_2, W=w) \sim N(1,1)$.
Without loss of generality, we fix $W=1$ and $y=1.808$. After some calculations, we obtain
\begin{eqnarray*}
P_{1\cdot01}=0.5,P_{1\cdot11}=0.8,P_{0\cdot11}=0.2,F_{01\cdot11}=0.7997,F_{11\cdot01}=0.1257,
F_{01\cdot01}=0.3743.
\end{eqnarray*}
Then, the right bound is $R_0(y,1|\pi_{2\cdot 1}^{\scriptstyle\max})=0.419$.
Consider the following linear programming problem subject to (\ref{linear programming 1}) given $W=1$ and $y=1.808$,
    \begin{equation*}
        \max \left\{F_{20\cdot 1}-F_{10\cdot 1}-0.419\times \left(F_{00\cdot 1}+F_{10\cdot 1}+F_{20\cdot 1}+F_{30\cdot 1}\right)\right\}.
    \end{equation*}
Solving the above linear program, we can obtain that the objective function's value is approximately $6.3832\times 10^{-6}$, which can be considered approximately equal to zero. The lower bound $L_0(y,w|\pi_{2\cdot w}^{\scriptstyle\max})$ can be illustrated by analogy. In fact, this holds for any grid point $y\in \mathcal{Y}$.  
\end{remark}

Most previous studies have usually focused on bounding the SACE defined in the always-survivor strata under the principal stratification framework proposed by \cite{2002Frangakis} because this allows the formulation of a well-defined causal effect on outcomes truncated by death. However, determining bounds on the causal effects in other strata might also yield useful information in applications. When studying the effects of job training on wages that discussed in Section~\ref{sec:real-data}, researchers are often interested in whether changes in employment status resulting from job training are accompanied by increases in potential outcomes. If $\Delta_1 >0$, it indicates that work experience has a positive effect in improving employment what policymakers might be concerned about. In some medical clinical experiments and socio-demographic surveys, the `harmed' strata may receive more attention. Here, we derive bounds on the ACEs in the `protected' and `harmed' strata. Given the covariate vector $W=w$, the conditional ACEs in the `protected' and `harmed' strata denoted by $\Delta_ {1\cdot w}$ and $\Delta_ {2\cdot w}$ respectively, are defined as follows:
\begin{equation}
    \Delta_ {k\cdot w} = E\{Y(1)-Y(0) \mid G=g_k,W=w\} , \quad \mbox{for} \,\, k\in\{1,2\}.
\end{equation}
The ACEs in these two strata  are denoted by $\Delta_ {1}$ and $\Delta_ {2}$ respectively:
\begin{equation}
\Delta_1={E\{\pi_{1 \cdot W}\Delta_{1,W}\}}/{E(\pi_{1 \cdot W})},\quad\quad \Delta_2={E\{\pi_{2 \cdot W}\Delta_{2,W}\}}/{E(\pi_{2 \cdot W})}.
\end{equation}

Using a method similar to that for treating $\mbox{SACE}_w$, we obtain
\begin{eqnarray}\label{linear_programming_delta1}
\begin{aligned}
\Delta_{1\cdot w }&= \int_{-\infty}^{+\infty}\Big \{F(y \mid Z=1, G=g_1, W=w) - F(y \mid Z=0, G=g_1, W=w) \Big\} dy,\\
&=\int_{-\infty}^{+\infty}  \frac {F_{21 \cdot w} -F_{11 \cdot w} }{F_{01 \cdot w} + F_{11 \cdot w} + F_{21 \cdot w} +F_{31 \cdot w}}  dy.
\end{aligned}
\end{eqnarray}
Likewise, for the `harmed' strata, we have
\begin{eqnarray}\label{delta2}
\begin{aligned}
\Delta_{2\cdot w }&=\int_{-\infty}^{+\infty}  \frac {F_{22 \cdot w} -F_{12 \cdot w} }{F_{02 \cdot w} + F_{12 \cdot w} + F_{22 \cdot w} +F_{32 \cdot w}}  dy.
\end{aligned}
\end{eqnarray}
Now, bounding the integrand term in (\ref{linear_programming_delta1}) and (\ref{delta2}) respectively for any fixed $y\in R$ subject to (\ref{linear programming 1}) leads to the following sharp analogue of Theorem \ref{thm-SACE} for $\Delta_{1\cdot w },\Delta_{1}, \Delta_{2\cdot w }$ and $\Delta_{2}.$ Let $\pi_{2\cdot w}^{\scriptstyle\min}=\max\{0,P_{1\cdot 0w}- P_{1\cdot 1w}\}$, which is the minimum value of $\pi_{2\cdot w}$. Define
\begin{eqnarray*}
    \begin{aligned}
    & L_1(y,w|\pi_{2\cdot w})= \max \left\{
\frac {-F_{11 \cdot 1w}}{P_{1\cdot 1w}-P_{1\cdot 0w} + \pi_{2\cdot w}}, 
      -1
\right\}, \\
&R_1(y,w|\pi_{2\cdot w})= \min \left\{
\frac {F_{01 \cdot 1w}}{P_{1\cdot 1w}-P_{1\cdot 0w} + \pi_{2\cdot w}},
1
\right\},
\\
 &L_1=\frac {E\Big\{(P_{1\cdot 1W}-P_{1\cdot 0W} + \pi_{2\cdot W}^{\scriptstyle\min})\int_{-\infty}^{+\infty}L_1(y,W|\pi_{2\cdot W}^{\scriptstyle\min})dy\Big\}}{E(P_{1\cdot 1W}-P_{1\cdot 0W} + \pi_{2\cdot w}^{\scriptstyle\min})},
        \\
        &R_1=\frac {E\Big\{(P_{1\cdot 1W}-P_{1\cdot 0W} + \pi_{2\cdot W}^{\scriptstyle\min})\int_{-\infty}^{+\infty}R_1(y,W|\pi_{2\cdot W}^{\scriptstyle\min})dy\Big\}}{E(P_{1\cdot 1W}-P_{1\cdot 0W} + \pi_{2\cdot W}^{\scriptstyle\min})},
    \end{aligned}
\end{eqnarray*}
and
\begin{eqnarray*}
    \begin{aligned}
&R_2(y,w|\pi_{2\cdot w})= \min \left\{
\frac {F_{11 \cdot 0w}}{ \pi_{2\cdot w}},1 
\right\}, \quad\quad\quad L_2(y,w|\pi_{2\cdot w})= \max \left\{
\frac {-F_{01 \cdot 0w}}{ \pi_{2\cdot w}},-1
\right\},\\
 &L_2=\frac {E\Big\{\pi_{2\cdot W}^{\scriptstyle\min}\int_{-\infty}^{+\infty}L_2(y,W|\pi_{2\cdot W}^{\scriptstyle\min})dy\Big\}}{E(\pi_{2\cdot W}^{\scriptstyle\min})},\quad R_2=\frac {E\Big\{\pi_{2\cdot W}^{\scriptstyle\min}\int_{-\infty}^{+\infty}R_2(y,W|\pi_{2\cdot W}^{\scriptstyle\min})dy\Big\}}{E(\pi_{2\cdot W}^{\scriptstyle\min})}.
    \end{aligned}
\end{eqnarray*}

\begin{theorem}
     \label{thm-Delta12}Under Assumptions 1-2, given covariates $W=w,$ the $\Delta_{1\cdot w }$ and $\Delta_{2\cdot w }$ are bounded as
\begin{eqnarray}\left\{
    \begin{aligned}
        &  \int_{-\infty}^{+\infty}  L_1(y,w|\pi_{2\cdot w}^{\scriptstyle\min})\, dy\leq \Delta_{1\cdot w } \leq \int_{-\infty}^{+\infty}  R_1(y,w|\pi_{2\cdot w}^{\scriptstyle\min})\, dy,\\
        & \int_{-\infty}^{+\infty}  L_2(y,w|\pi_{2\cdot w}^{\scriptstyle\min})\, dy\leq \Delta_{2\cdot w } \leq \int_{-\infty}^{+\infty}  R_2(y,w|\pi_{2\cdot w}^{\scriptstyle\min})\, dy,
    \end{aligned}\right .
\end{eqnarray}
and the $\Delta_{1}$ and $\Delta_{2}$ are bounded as
\begin{eqnarray}
\begin{aligned}
L_1\leq \Delta_1 \leq R_1,\quad
L_2\leq \Delta_2 \leq R_2.
\end{aligned}
\end{eqnarray}
\end{theorem}
As the proof of this result closely follows the proof of Theorem~\ref{thm-SACE}, we will only sketch the necessary modiﬁcations to the proof in Appendix \ref{Appendix: thm-Delta12}. Compared with the results given by \citet{2015Huber} which restricted outcome to be bounded, the above resulting bounds allow for the setting of square-integrable outcome. 

\section{Nonparametric bounds with monotonicity and/or 
stochastic dominance} 

The bounds derived in the previous section may be too wide to be useful in applications, for example, when judging the sign of the causal effect. To obtain much tighter bounds on the causal effect, we require the monotonicity assumption and/or the stochastic dominance assumption.

\subsection{Monotonicity}
It is expected that by imposing the assumption of monotonicity of the survival status under treatment in addition to Assumptions 1 and 2, the width of the bounds could be improved. Monotonicity of the survival status implies that the treatment does not cause death compared to the control, meaning that if the subject dies before the outcome can be measured under treatment, then the subject would also have died before the outcome could be measured in the control case.
\begin{assumption}[Monotonicity]\label{assump3} $\P\{S_i(1) \geq\ S_i(0)\}=1.$ 
\end{assumption}
This assumption is reasonable for many studies. One type of monotonicity assumption is that if an adverse event occurs for a patient assigned to a placebo, then that adverse event would also occur if this same patient were assigned to the experimental treatment. Another type of monotonicity assumption is that people who receive training are more likely to find new jobs in the NSW trials. This assumption rules out the existence of a `harmed' subpopulation (strata $g_2$), that is, $\P(G = g_2) = 0$, which makes it possible to identify each proportion of the principal strata $\pi_{k \cdot w}$, $k\in \{0,1,3\}$ given covariate $W=w$. Under Assumptions 1-3, the term $\pi_{k\cdot w}$ can be rewritten as follows:
\begin{eqnarray}\label{4pi_0}
\left\{
\begin{aligned}
&\pi_{0\cdot w} =\P(S = 1 \mid Z=0, W=w),\\
&\pi_{1\cdot w}=\P(S = 1 \mid Z=1, W=w) - \P(S = 1 \mid Z=0, W=w),\\
&\pi_{3\cdot w} = \P(S = 0 \mid Z=1, W=w).
\end{aligned}
\right.
\end{eqnarray}

In fact, Assumption~\ref{assump3} implies that $\{S(0)=1\}$ is equivalent to $\{S(0)=1, S(1)=1\}$, that is, $S(0)=1$ is determined by strata $g_0$. As a result, the observed conditional mean outcome $\mu_{0,0,w} = E(Y \mid Z=0, G=g_0, W=w)=E(Y \mid Z=0, S=1, W=w)$ is identifiable. However, $\mu_{1,0,w}= E(Y \mid Z=1, G=g_0, W=w)$ is not identifiable from the observed data since both the `always-survivor' and `protected' strata are included in $S(1)=1$, which leads to, 
\begin{equation*}
F(y \mid Z = 1, S = 1,w) = \sum_{k=0, 1}\frac{\pi_{k \cdot w}}{\pi_{0 \cdot w}+\pi_{1 \cdot w}} \cdot F(y \mid Z = 1, G = g_k,w). 
\end{equation*}
As a result, $\mbox{SACE}_w=\mu_{1,0,w}-\mu_{0,0,w}$ is not identifiable. We shall use the following notation, 
\begin{eqnarray*}
\begin{aligned}
& L_{0,m}(y,w)=\max\left\{-\frac {F_{01 \cdot 0w}}{P_{1\cdot 0w}}, \frac {F_{11 \cdot 0w}-F_{11 \cdot 1w}}{P_{1\cdot 0w}}\right\},\\
&R_{0,m}(y,w)= \min\left\{\frac {F_{11 \cdot 0w}}{P_{1\cdot 0w}}, \frac {F_{01 \cdot1w}-F_{01 \cdot 0w}}{P_{1\cdot 0w}}\right\},\\
&L_{0,m}=\frac{E\Big\{P_{1\cdot 0W}\int_{-\infty}^{+\infty}  L_{0,m}(y,W)\, dy\Big\}}{E(P_{1\cdot 0W})},\,\,
R_{0,m}=\frac{E\Big\{P_{1\cdot 0W}\int_{-\infty}^{+\infty}  R_{0,m}(y,W)\, dy\Big\}}{E(P_{1\cdot 0W})}.
\end{aligned}
\end{eqnarray*}

\begin{theorem}
     \label{thm-SACE-m}Under Assumptions \ref{assu1}-\ref{assump3}, the conditional $\mbox{SACE}_{w }$ and SACE are bounded as
\begin{equation}\label{bound_11_w_m}
 \int_{-\infty}^{+\infty}  L_{0,m}(y,w)\, dy\leq \mbox{SACE}_{w } \leq \int_{-\infty}^{+\infty}  R_{0,m}(y,w)\, dy,
\end{equation} 
and
\begin{equation}\label{bound_11_m}
L_{0,m} \leq \mbox{SACE} \leq R_{0,m}.
\end{equation}
\end{theorem}
The proof of Theorem~\ref{thm-SACE-m} is completed in Appendix~\ref{Appendix: thm-SACE-m}. We can compare the bounds given in (\ref{bound_11_m}) with those in the literature in the context in which the model includes or excludes covariates. If $Y$ is binary and the model excludes covariates, (\ref{bound_11_m}) reduces to the bounds of \citet{2015Shan}, while if $Y$ is nonnegative and the model includes covariates, (\ref{bound_11_m}) coincides with the bounds on $\beta$ in Theorem~1 of \citet{2019Kennedy}.

Next, we will deduce bounds on the ACE only in the `protected' strata; this is because the `harmed' strata no longer exists under Assumption 3, while the `doomed' strata is unavailable since we cannot observe any related information. Under Assumption~\ref{assump3}, the proportion of the `protected' strata, $\pi_{1\cdot w}$, is point-identified, as illustrated in (\ref{4pi_0}). Let
\begin{eqnarray*}
\begin{aligned}
&R_{1,m}(y,w)= \min \left\{
\frac {F_{01 \cdot 1w}}{P_{1\cdot 1w}-P_{1\cdot 0w} },
1
\right\},\quad L_{1,m}(y,w)= \max \left\{
\frac {-F_{11 \cdot 1w}}{P_{1\cdot 1w}-P_{1\cdot 0w} }, 
      -1
\right\},\\
&L_{1,m}=\frac{E\Big\{(P_{1\cdot 1W}-P_{1\cdot 0W})\int_{-\infty}^{+\infty}  L_{1,m}(y,W)\, dy\Big\}}{E(P_{1\cdot 1W}-P_{1\cdot 0W})},\\
&R_{1,m}=\frac{E\Big\{(P_{1\cdot 1W}-P_{1\cdot 0W})\int_{-\infty}^{+\infty}  R_{1,m}(y,W)\, dy\Big\}}{E(P_{1\cdot 1W}-P_{1\cdot 0W})}.
\end{aligned}
\end{eqnarray*} 
We can state the result of bounds on the ACE in the `protected' strata.
\begin{theorem}
     \label{thm-Delta1-m}Under Assumptions \ref{assu1},\ref{assu2} and \ref{assump3}, the conditional $\Delta_{1\cdot w }$ and $\Delta_1$ are bounded as
\begin{equation}\label{bound_10_w_m}
 \int_{-\infty}^{+\infty}  L_{1,m}(y,w)\, dy\leq \Delta_{1\cdot w} \leq \int_{-\infty}^{+\infty}  R_{1,m}(y,w)\, dy,
\end{equation} 
and
\begin{equation}\label{bound_10_m}
L_{1,m} \leq \Delta_{1} \leq R_{1,m}.
\end{equation}
\end{theorem}
We sketch the proof of Theorem~\ref{thm-Delta1-m} in Appendix \ref{Apppendix:thm-Delta1-m}. 
In the absence of Assumption~\ref{assump3},  Theorem~\ref{thm-Delta12} gives a bound on $\Delta_{1\cdot w}$.  The main point is that the denominator of this bound makes sense when $\pi_{2\cdot w}^{min}=0$ which implies that $P_{1\cdot 1w} > P_{1\cdot 0w}$. Moreover, when $\pi_{2\cdot w}^{min}=0$, (\ref{bound_10_w_m}) coincides with the bounds under Assumption~\ref{assump3} given in Theorem~\ref{thm-Delta1-m}. For the `protected' strata, the monotonicity assumption (Assumption~\ref{assump3}) does not tighten the bounds further. Similar results have been obtained in the literature. \citet{1997Balke}, \citet{2001Heckman_Vytlacil} and \citet{2021Kitagawa} showed that under Assumption~\ref{assump3}, their bounds on the ACE in the entire population coincided with the bounds of \citet{1990Manski}, who invoked only mean independence in the entire population. This also reveals that Assumption~3 does not provide any additional identifying power for the ACE when it is satisfied.

In general, suppose that the monotonicity assumption (Assumption~\ref{assump3}) allows us to avoid investigating the identifiability of the principal stratum proportions $\pi_{k\cdot w}$, $k=0,1.$ Appendix \ref{Appendix: thm-SACE-m} and \ref{Apppendix:thm-Delta1-m} show that this reduces the number of variables and avoids the parameter in linear programming problems (\ref{fracPf0_mon}) and (\ref{fracPf1_mon}). Furthermore, the application example on a real dataset presented in Section~\ref{sec:real-data} will show that the monotonicity assumption helps to tighten the bounds on the $\mbox{SACE}$.

\subsection{Stochastic dominance}
\label{section:without M}
The stochastic dominance assumption, which means that one probability distribution always coincides with or lies to the right of another probability distribution, has been used by \citet{2008Zhang}, \citet{2015Huber}, and \citet{2007Blundell}. It is stated formally as follows:
\begin{assumption}\label{assu4}
[Stochastic Dominance]
\begin{equation*}
          \P(Y_i(1) \leq y \mid G=g_0,W=w) \leq \P(Y_i(1) \leq y \mid G=g_1,W=w), \,\,\mbox{for}\,\, y\in R,
\end{equation*}
and
\begin{equation*}
          \P(Y_i(0) \leq y \mid G=g_0,W=w) \leq \P(Y_i(0) \leq y \mid G=g_2,W=w),  \,\,\mbox{for}\,\, y\in R.
\end{equation*}
\end{assumption}
This assumption means that the probability distribution of $Y(1)$ among the `always-survivor' subpopulation coincides with or lies to the right of $Y(1)$ among the `protected' subpopulation and that the probability distribution of $Y(0)$ among the `always-survivor' subpopulation coincides with or lies to the right of $Y(0)$ among the `harmed' subpopulation. When assessing the return of job training, this assumption means that the potential wages observed are always at least as high as those of other groups. This may be because members of the `always-survivor' subpopulation will be hired regardless of training, since they are likely more motivated and/or capable than the rest of the population. \citet*{2008Zhang} argued that ability tends to be positively correlated with wages, so Assumption~\ref{assu4} appears to be plausible. It is known (see, e.g.,\citealp{2003Muller}) that Assumption~\ref{assu4} implies that
\begin{eqnarray*}
    \left\{\begin{aligned}        
          & E\{Y(1) \mid G=g_0,W=w) \geq  E(Y(1) \mid G=g_1,W=w\},\\
          &  E\{Y(0) \mid G=g_0,W=w) \geq  E(Y(0) \mid G=g_2,W=w\}.
    \end{aligned}
    \right.
\end{eqnarray*}
The mean earnings of the members of the `always-survivor' stratum in the treatment arm are greater than or equal to the mean earnings of the members of the `protected' stratum in the treatment arm. Similarly, the mean earnings of the members of the `always-survivor' stratum in the control arm are greater than or equal to the mean earnings of the members of the `harmed' stratum in the control arm. 

While considering further Assumption~\ref{assu4}, one can immediately obtain bounds on the ACEs in three strata by transforming this assumption to the resulting constraints on $F_{ij\cdot w}$. The next theorem is the main result of this section. We need to introduce a bit more notation. Define
\begin{eqnarray*}
    \begin{aligned}
        & L_{0,\mbox{sd}}(y,w\mid \pi_{2\cdot w})= \max \left\{\frac {-F_{11 \cdot 1w}}{P_{1\cdot 1w}}, \frac {F_{01 \cdot 1w}}{P_{1\cdot 1w}} - \frac {F_{01 \cdot 0w}}{P_{1\cdot 0w}-\pi_{2\cdot w}}
\right\},\\
        & L_{0,\mbox{sd}}=\frac{E\left\{(P_{1\cdot 0W}-\pi_{2\cdot W}^{\scriptstyle\max}) \int_{-\infty}^{+\infty} L_{0,\mbox{sd}}(y,W\mid \pi_{2\cdot W}^{\scriptstyle\max})dy\right\}}{E(P_{1\cdot 0W}-\pi_{2\cdot W}^{\scriptstyle\max})}, \\
        & R_{0,\mbox{sd}}(y,w\mid \pi_{2\cdot w}) = \min \left\{
\frac {F_{11 \cdot 0w}}{P_{1\cdot 0w}}, \frac {-F_{01 \cdot 0w}}{P_{1\cdot 0w}} + \frac {F_{01 \cdot 1w}}{P_{1\cdot 0w}-\pi_{2\cdot w}} \right\},\\
        & R_{0,\mbox{sd}}=\frac{E\left\{ (P_{1\cdot 0W}-\pi_{2\cdot w}^{\scriptstyle\max})\int_{-\infty}^{+\infty}R_{0,\mbox{sd}}(y,W\mid \pi_{2\cdot W}^{\scriptstyle\max})dy\right\}}{E(P_{1\cdot 0W}-\pi_{2\cdot W}^{\scriptstyle\max})},
\end{aligned}
\end{eqnarray*}
and
\begin{eqnarray*}
\begin{aligned}
&L_{1,\mbox{sd}}(y,w|\pi_{2\cdot w})= \max \left\{
\frac {-F_{11 \cdot 1w}}{P_{1\cdot 1w}-P_{1\cdot 0w}+\pi_{2\cdot w}}, 
       -1
\right\},\,\,\,\,R_{1,\mbox{sd}}(y,w|\pi_{2\cdot w})= \frac {F_{01 \cdot 1w}}{P_{1\cdot 1w}},\\
& L_{1,\mbox{sd}} = \frac{E\{(P_{1\cdot 1W}-P_{1\cdot 0W}+\pi_{2\cdot W}^{\scriptstyle\min})\int_{-\infty}^\infty L_{1,\mbox{sd}}(y,W|\pi_{2\cdot W}^{\scriptstyle\min})\,dy\}}{E(P_{1\cdot 1W}-P_{1\cdot 0W}+\pi_{2\cdot W}^{\scriptstyle\min})} ,\\
& R_{1,\mbox{sd}} = \frac{E\{(P_{1\cdot 1W}-P_{1\cdot 0W}+\pi_{2\cdot W}^{\scriptstyle\min})\int_{-\infty}^\infty R_{1,\mbox{sd}}(y,W|\pi_{2\cdot W}^{\scriptstyle\min})\,dy\}}{E(P_{1\cdot 1W}-P_{1\cdot 0W}+\pi_{2\cdot W}^{\scriptstyle\min})},\\
&L_{2,\mbox{sd}}(y,w|\pi_{2\cdot w})= 
\frac {-F_{01 \cdot 0w}}{P_{1\cdot 0w}},\hspace{2.2cm}
R_{2,\mbox{sd}}(y,w|\pi_{2\cdot w})= \min \left\{
\frac {F_{11 \cdot 0w}}{\pi_{2\cdot w}}, 1
\right\},\\
& L_{2,\mbox{sd}} = \frac{E\{\pi_{2\cdot W}^{\scriptstyle\min}\int_{-\infty}^\infty L_{2,\mbox{sd}}(y,W)\,dy\}}{E(\pi_{2\cdot W}^{\scriptstyle\min})},\quad
R_{2,\mbox{sd}} = \frac{E\{\pi_{2\cdot W}^{\scriptstyle\min}\int_{-\infty}^\infty R_{2,\mbox{sd}}(y,W|\pi_{2\cdot W}^{\scriptstyle\min})\,dy\}}{E(\pi_{2\cdot W}^{\scriptstyle\min})}. 
\end{aligned}
\end{eqnarray*}

\begin{theorem}
     \label{thm-Deltas-sd}Under Assumptions \ref{assu1}-\ref{assu2} and \ref{assu4}, for $k=1,2,$ the conditional $\mbox{SACE}_w,\Delta_{k\cdot w }$ and $\mbox{SACE},\Delta_{k}$ are bounded as
\begin{eqnarray}
    \left\{\begin{aligned} 
        &\int_{-\infty}^{\infty}L_{0,\mbox{sd}}(y,w\mid \pi_{2\cdot w}^{\scriptstyle\max})\,dy \leq \mbox{SACE}_w \leq \int_{-\infty}^{\infty}R_{0,\mbox{sd}}(y,w\mid \pi_{2\cdot w}^{\scriptstyle\max})\,dy,\\
       & \int_{-\infty}^{\infty} L_{k,\mbox{sd}}(y,w\mid \pi_{2\cdot w}^{\scriptstyle\min})\,dy\leq \Delta_{k\cdot w}\leq \int_{-\infty}^{\infty}R_{k,\mbox{sd}}(y,w\mid \pi_{2\cdot w}^{\scriptstyle\min})\,dy,         
    \end{aligned}
    \right.
\end{eqnarray}
and
\begin{eqnarray}
    \begin{aligned} 
        L_{0,\mbox{sd}}\leq \mbox{SACE}\leq R_{0,\mbox{sd}},
        \quad     L_{k,\mbox{sd}}\leq \Delta_{k}\leq R_{k,\mbox{sd}}.   
    \end{aligned}
\end{eqnarray}
\end{theorem}
We sketch the proof of Theorem~\ref{thm-Deltas-sd} in Appendix \ref{Appendix: thm-Deltas-sd}. 
Note that the stochastic dominance assumption effectively tightens the bounds on the $\mbox{SACE}$, but only the upper bound for the `protected' stratum and the lower bound for the `harmed' stratum. That is, $L_{1,\mbox{sd}}=L_{1}$ and $R_{2,\mbox{sd}}=R_{2}.$

\subsection{Monotonicity and stochastic dominance}
\label{section:with all}
Below, we derive the bounds when both the monotonicity and stochastic dominance assumptions hold. Since the `harmed' stratum is excluded under the monotonicity assumption, we consider only the bounds in the `always-survivor' and `protected' strata. Then, the bounds on the SACE and $\Delta_1$ under Assumptions \ref{assu1}-\ref{assu4} can be obtained as follows:
\begin{theorem}
     \label{thm-Deltas-m-sd}Under Assumptions \ref{assu1}-\ref{assu4}, the conditional $\mbox{SACE}_w,\Delta_{1\cdot w }$ and $\mbox{SACE},\Delta_{1}$ are bounded as
\begin{eqnarray}
    \left\{\begin{aligned} 
        &    \int_{-\infty}^{\infty}L_{0,\mbox{m},\mbox{sd}}(y,w)\,dy \leq \mbox{SACE}_w \leq \int_{-\infty}^{\infty}R_{0,\mbox{m},\mbox{sd}}(y,w)\,dy,\\
       &     \int_{-\infty}^{\infty}L_{1,\mbox{m},\mbox{sd}}(y,w)\,dy \leq \Delta_{1\cdot w} \leq \int_{-\infty}^{\infty}R_{1,\mbox{m},\mbox{sd}}(y,w)\,dy,      
    \end{aligned}
    \right.
\end{eqnarray}
and
\begin{eqnarray}
    \begin{aligned} 
  L_{0,\mbox{m},\mbox{sd}}\leq \mbox{SACE}\leq  R_{0,\mbox{m},\mbox{sd}},
        \quad       L_{1,\mbox{m},\mbox{sd}}\leq \Delta_1\leq  R_{1,\mbox{m},\mbox{sd}}, 
    \end{aligned}
\end{eqnarray}
where
\begin{eqnarray*}
    \begin{aligned}
        & L_{0,\mbox{m,sd}}(y,w)= \max \left\{\frac {F_{01 \cdot 1w}}{P_{1 \cdot 1w}}-\frac {F_{01 \cdot 0w}}{P_{1\cdot 0w}}, \frac {F_{11 \cdot 0w}-F_{11 \cdot 1w}}{P_{1\cdot 0w}}\right\},\\
        & L_{0,\mbox{m,sd}}=\frac{E\left\{P_{1\cdot 0W}\int_{-\infty}^{+\infty} L_{0,\mbox{m,sd}}(y,W)dy\right\}}{E(P_{1\cdot 0W})}, \\
        & R_{0,\mbox{m,sd}}(y,w) = \min\left\{\frac {F_{11 \cdot 0w}}{P_{1\cdot 0w}}, \frac {F_{01 \cdot1w}-F_{01 \cdot 0w}}{P_{1\cdot 0w}}\right\},\\
        & R_{0,\mbox{m,sd}}=\frac{E\left\{ P_{1\cdot 0W}\int_{-\infty}^{+\infty}R_{0,\mbox{m,sd}}(y,W)dy\right\}}{E(P_{1\cdot 0W})}\\
         & L_{1,\mbox{m,sd}}(y,w)= \max \left\{
\frac {-F_{11 \cdot 1w}}{P_{1\cdot 1w}-P_{1\cdot 0w}}, 
       -1
\right\}, \,\,R_{1,\mbox{m,sd}}(y,w) = 
\frac {F_{01 \cdot 1w}}{P_{1\cdot 1w}},\\
&L_{1,\mbox{m,sd}}=\frac{E\left\{(P_{1\cdot 1W}-P_{1\cdot 0W})\int_{-\infty}^\infty L_{1,\mbox{m,sd}}(y,W)\,dy\right\}}{E(P_{1\cdot 1W}-P_{1\cdot 0W})},\\
&R_{1,\mbox{m,sd}}=\frac{E\left\{(P_{1\cdot 1W}-P_{1\cdot 0W})\int_{-\infty}^\infty R_{1,\mbox{m,sd}}(y,W)\,dy\right\}}{E(P_{1\cdot 1W}-P_{1\cdot 0W})}.
    \end{aligned}
\end{eqnarray*}
\end{theorem}
The detailed proof is given in Appendix \ref{Appendix: thm-Deltas-m-sd}.
The above results show that the bounds become tighter when both Assumption~\ref{assump3} and Assumption~\ref{assu4} are invoked. For the `always-survivor' stratum, the lower bounds on both $\mbox{SACE}_w$ and the $\mbox{SACE}$ are tightened. However, invoking both assumptions does not provide any additional identifying power for the ACE in the `protected' stratum, so these bounds coincide with those given in Theorem~\ref{thm-Deltas-sd} for the case in which only the stochastic dominance assumption is considered.

\section{Estimation}
In this section, we provide estimators of the bounds derived under the various assumptions based on the method of moments. Suppose that we observe an independent and identically distributed sample $(X_i,Z_i,S_i,Y_i)$, $1\leq i\leq n.$ We define the following estimators,
\begin{eqnarray*}
    \begin{aligned}
       & \hat{P}_{1 \cdot zw} = \sum_{i=1}^n S_iI(Z_i=z)K\{(W_i-w)/h\}\Big/
       \sum_{i=1}^n I(Z_i=z)K\{(W_i-w)/h\},\\
       & \hat{P}_{0 \cdot zw} = 1- \hat{P}_{1 \cdot zw},\\
       & \hat{F}_{11\cdot zw}(y) =  \sum_{i=1}^n I(Y_i\leq y)S_iI(Z_i=z)K\{(W_i-w)/h\}\Big/
       \sum_{i=1}^n I(Z_i=z)K\{(W_i-w)/h\},\\
       & \hat{F}_{01\cdot zw}(y) =  \hat{P}_{1 \cdot zw}-\hat{F}_{11\cdot zw}(y),
    \end{aligned}
\end{eqnarray*}
where $I(\cdot)$ is an indicator function and when $W_i$ is a discrete random variable, $K(\cdot)$ is also an indicator function that takes a value of $1$ if $W_i=w$, while when $W_i$ is a continuous covariate vector, $K(\cdot)$ is a kernel function, where the popular kernels include the normal kernel $K(t)=\exp(-t^2/2)/\sqrt{2\pi}$ and the Epanechnikov kernel $K(t)=3(1-t^2)I(|t|<1)/4$. In practice, it can be implemented using the R function `npcdist' from the `np' package , which is based on the work of \citet{2008Li_Racine}, who employed `generalized product kernels' that admit a mix of continuous and discrete data types. The estimators of various bounds given $w$ can be obtained by plugging in these expressions instead of the corresponding population parameters. The estimators of bounds related to the expectation of $W$ can be established by taking the mean over $\{W_i,1\leq i\leq n\}$. Following \citet{2009Lee}, $\sqrt{n}$-consistency and asymptotic normality of these estimators can be derived; we omit any further discussion of this here. We need to deal with the numerical evaluation (approximation) of definite integrals with respect to $y$ such as $\int_{-\infty}^{+\infty}  L_1(y,w|\pi_{2\cdot w}^{\scriptstyle\min})\, dy, \int_{-\infty}^{+\infty}  R_1(y,w|\pi_{2\cdot w}^{\scriptstyle\min})\, dy$. Consider a sufficiently large positive $M$. One way to proceed is to equally divide the range $[-M, M]$ into many small
intervals, say $m$ intervals of width $h = 2M/m$, which we call the step size. For each interval, we evaluate the integrand value and take that value to represent that interval. The other way to approach the problem is to calculate the area covered by the integrand function. Then, we can sum the contributions from each interval and consider that the summation should approximate the integral. Among the alternative methods for approximating an integral, Monte Carlo integration is one of the most widely used. In its simplest form, Monte Carlo integration approximates an integral by calculating the average for $N$ random samples chosen uniformly at random in $[-M,M]$. Furthermore, various techniques such as importance sampling and stratified sampling have been proposed to reduce the variance of this method and enhance its convergence. The convergence rates of these integral approximation methods can be sufficiently fast as to have no effect on the $\sqrt{n}$-consistency and asymptotic normality of the proposed estimators.

\section{Simulation studies}
\label{sec:simu}

In this section, we report some simulation studies conducted to evaluate the finite-sample performance of the proposed methods in various cases with and without the monotonicity and stochastic dominance assumptions. We also consider cases of both a discrete and a continuous response $Y$. In the following subsections, we generate a sample of size $1000$ from each of several frequently used continuous and discrete distributions, and the averages of the estimated lower and upper bounds are computed over $1000$ replications. We define $F_{z\mid k,w}=\P(Y_i(z) \leq y \mid G=g_k,W=w)$, ${\mu}_{z,k,w}=E(Y|Z=z,G=g_k,W=w)$, $\sigma_{z,k,w}^2=\mbox{Var}(Y|Z=z,G=g_k,W=w)$ and $\pi_{k\cdot w} = \P(G = g_k \mid W=w)$. The considered distributions are listed as follows:
\begin{itemize}    
    \item {\bf Normal Distribution.} The outcome $Y$ is drawn from the following normal distribution conditional on $Z$, $G$ and $W$: $Y \mid (Z=1, G=g_0, W=w) \sim N(\mu_{1,0,w},\sigma_{1,0,w}^2)$, \,$Y \mid (Z=1, G=g_1, W=w) \sim  N(\mu_{1,1,w},\sigma_{1,1,w}^2)$,\,$Y \mid (Z=0, G=g_0, W=w) \sim  N(\mu_{0,0,w},\sigma_{0,0,w}^2)$ and $Y \mid (Z=0, G=g_2,W=w) \sim N(\mu_{0,2,w},\sigma_{0,2,w}^2)$.
    \item {\bf Normal Mixture Distribution.} The outcome $Y$ is generated as follows: 
    $Y \mid (Z=1, G=g_0,W=w) \sim \omega_{1,0} N(\mu_{1,0,w}^{(1)},\sigma^{2(1)}_{1,0,w}) + (1-\omega_{1,0})N(\mu_{1,0,w}^{(2)},\sigma^{2(2)}_{1,0,w})$,\,$Y \mid (Z=1, G=g_1.W=w) \sim  \omega_{1,1}N(\mu_{1,1,w}^{(1)},\sigma^{2(1)}_{1,1,w}) + (1-\omega_{1,1}) N(\mu_{1,1,w}^{(2)},\sigma^{2(2)}_{1,1,w})$,\,$ Y \mid (Z=0, G=g_0, W=w) \sim \omega_{0,0} N(\mu_{0,0,w}^{(1)},\sigma^{2(1)}_{0,0,w}) + (1-\omega_{0,0}) N(\mu_{0,0,w}^{(2)},\sigma^{2(2)}_{0,0,w})$ and $ Y \mid (Z=0, G=g_2, W=w) \sim \omega_{0,2} N(\mu_{0,2,w}^{(1)},\sigma^{2(1)}_{0,2,w}) + (1-\omega_{0,2}) N(\mu_{0,2,w}^{(2)},\sigma^{2(2)}_{0,2,w})$.
   \item {\bf Binary Distribution.} A binary $Y$ follows a Bernoulli distribution conditional on $Z$, $G$ and $W$ where $Y \mid (Z=1, G=g_0, W=w) \sim B(1, \mu_{1,0,w})$, $Y \mid (Z=1, G=g_1, W=w) \sim B(1, \mu_{1,1,w}), Y \mid (Z=0, G=g_0, W=w) \sim B(1, \mu_{0,0,w})$.
    \item {\bf Poisson Distribution.} The outcome $Y$ is generated from the following Poisson distribution conditional on $Z$, $G$ and $W$: $Y \mid (Z=1, G=g_0, W=w) \sim \mbox{Pois}(\mu_{1,0,w})$,\,$Y \mid (Z=1, G=g_1, W=w) \sim  \mbox{Pois}(\mu_{1,1,w})$,\,$ Y \mid (Z=0, G=g_0, W=w) \sim  \mbox{Pois}(\mu_{0,0,w})$ and $Y \mid (Z=0, G=g_2, W=w) \sim \mbox{Pois}(\mu_{0,2,w})$.
    
\end{itemize}
The above terms with a subscript $w$ (for example, $\mu_{1,0,w}$) indicate the existence of covariates, while those without a subscript $w$ (for example, $\mu_{1,0}$) indicate that there are no covariates. 

Due to space limitations, we present simulation results of only two cases: the case without Assumptions~\ref{assump3} and \ref{assu4}, and the case with Assumptions~\ref{assump3} and \ref{assu4}. The other two cases with Assumption~\ref{assump3} or  Assumption~\ref{assu4} are relegated in \ref{Appendix: simu}.

\subsection{The case without Assumptions~\ref{assump3} and \ref{assu4}}
\label{no34}
When both Assumptions~\ref{assump3} and \ref{assu4} are not true, the approach described in Section \ref{section:withoutM&SD} is used to derive the bounds. Here, we assess how sensitive the proposed method is to the departure from Assumptions~\ref{assump3} and \ref{assu4}. To run the simulation, we first the following parameter values: $\pi_{1} = 0.3$, $\pi_{3} = 0.1$ and $\pi_{0} = 0.6 -\pi_{2}$ where $\pi_{2}$ is varied from $0.05$ to $0.25$. Here $\pi_2\neq 0$ implies that Assumption~\ref{assump3} does not hold. Then, we consider four cases of violation of Assumption~\ref{assu4} where either 1) $\mu_{1,0}<\mu_{1,1}$ and $\mu_{0,0}<\mu_{0,2}$ or 2) $\sigma_{1,0}^2<\sigma_{1,1}^2$ and $\sigma_{0,0}^2<\sigma_{0,2}^2$ under the assumption that $\mu_{1,0}=\mu_{1,1}$ and $\mu_{0,0}=\mu_{0,2}.$ The true value of the SACE in the first two cases is 4, that in the third case is $-4$, and that in the fourth case is $-5$. The simulated data are generated as follows:

\begin{itemize}    
    \item {\bf The case of a normal $Y$ with $\mu_{1,0}<{\mu}_{1,1}$ and ${\mu}_{0,0}<{\mu}_{0,2}$.} The parameters are $\mu_{1,0}=7,\mu_{1,1}=8,\mu_{0,0}=3,\mu_{0,2}=4$ and $,\sigma_{1,0}^2=\sigma_{1,1}^2=1/2,\sigma_{0,0}^2=\sigma_{0,2}^2=1.$
    \item {\bf The case of a normal $Y$ with $\sigma_{1,0}^2<\sigma_{1,1}^2$ and $\sigma_{0,0}^2<\sigma_{0,2}^2$.} The parameters are $\sigma_{1,0}^2=1/2,\sigma_{1,1}^2=1,\sigma_{0,0}^2=1/2,\sigma_{0,2}^2=1,$ $\mu_{1,0}=\mu_{1,1}=7,$ and $\mu_{0,0}=\mu_{0,2}=3.$    
    \item {\bf The case of a normal mixture $Y$ with ${\mu}_{1,0}<{\mu}_{1,1}$ and ${\mu}_{0,0}<{\mu}_{0,2}$.} The parameters are
    $\omega_{1,0}=0.5,\omega_{1,1}=0.4,\mu_{1,0}^{(1)}=2,\mu_{1,0}^{(2)}=4,\mu_{1,1}^{(1)}=2,\mu_{1,1}^{(2)}=5$ with $\sigma^{2(1)}_{1,0}=\sigma^{2(2)}_{1,0}=\sigma^{2(1)}_{1,1}=\sigma^{2(2)}_{1,1}=1$ and $\omega_{0,0}=0.5,\omega_{0,2}=0.5,\mu_{0,0}^{(1)}=6,\mu_{0,0}^{(2)}=8,\mu_{0,2}^{(1)}=6,\mu_{0,2}^{(2)}=7$ with $\sigma^{2(1)}_{0,0}=\sigma^{2(2)}_{0,0}=\sigma^{2(1)}_{0,2}=\sigma^{2(2)}_{0,2}=1.$ 
    \item {\bf The case of a Poisson $Y$ with ${\mu}_{1,0}<{\mu}_{1,1}$ and ${\mu}_{0,0}<{\mu}_{0,2}$.} The parameters are 
    $\mu_{1,0}=2,\mu_{1,1}=1,\mu_{0,0}=7,\mu_{0,2}=8.$ 
\end{itemize}
For illustration the distribution functions given $G$ in the various cases are shown in Figure \ref{Vio_A4}, which shows that Assumption~4 is violated. 

\begin{figure}[htbp]
  \centering
  \includegraphics[width=0.8\textwidth]{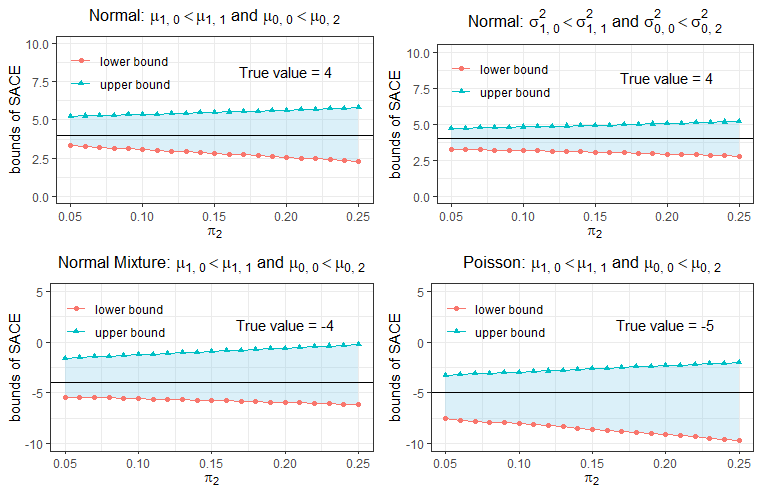}
  \caption{Bounds on the SACE without Assumptions 3 and 4.}\label{fig:simu_12}
\end{figure}

Figure \ref{fig:simu_12} illustrates that when both Assumptions~\ref{assump3} and \ref{assu4} do not hold, the bounds on the SACE contain the true value, and their range varies as $\pi_2$ varies over the interval $[0.05,0.25]$. All bounds never include zero when $\pi_2$ lies within its domain, which implies that the proposed bounds provide valid information.



\subsection{The case with Assumptions 3 and 4}
Further, we evaluate the performance of the proposed method in Section \ref{section:with all}, for the case in which all our assumptions hold. Here, we examine simulations in which the availability of a covariate $W$ may lead to narrower bounds, where $W\mid (Z=z,g_k) \sim B(1,0.4), k=0,1$. Let $\pi_{0\cdot w}=0.5,\pi_{1\cdot w}=0.4$ and 
$\pi_{3\cdot w}=0.1$. Simulations are carried out for the following cases, where $Y$ is drawn from a normal, normal mixture, binary or poisson distribution.
\begin{itemize}    
    \item {\bf The case of a normal $Y$.} The parameters are as follows: given $W=1$, $\mu_{1,0,1}=2$, $\mu_{1,1,1}=1$, $\mu_{0,0,1}=6$ with $\sigma_{1,0,1}^2=\sigma_{1,1,1}^2=\sigma_{0,0,1}^2=1$; given $W=0$, $\mu_{1,0,0}=5,\mu_{1,1,0}=4,\mu_{0,0,0}=3$ with $\sigma_{1,0,0}^2=\sigma_{1,1,0}^2=\sigma_{0,0,0}^2=1$.
    \item  {\bf The case of a normal mixture $Y$.} The parameters are: $\omega_{1,0}=\omega_{1,1}=\omega_{0,0}=0.4,$, and given $W=1$, $\mu_{1,0,1}^{(1)}=6, \mu_{1,0,1}^{(2)}=5, \mu_{1,1,1}^{(1)}=6, \mu_{1,1,1}^{(2)}=4, \mu_{0,0,1}^{(1)}=1, \mu_{0,0,1}^{(2)}=2$ with $\sigma^{2(1)}_{1,0,1}=\sigma^{2(2)}_{1,0,1}=\sigma^{2(1)}_{1,1,1}=\sigma^{2(2)}_{1,1,1}=\sigma^{2(1)}_{0,0,1}=\sigma^{2(2)}_{0,0,1}=1$, given $W=0$,  $\mu_{1,0,0}^{(1)}=5, \mu_{1,0,0}^{(2)}=2, \mu_{1,1,0}^{(1)}=4, \mu_{1,1,0}^{(2)}=1, \mu_{0,0,0}^{(1)}=3, \mu_{0,0,0}^{(2)}=6$ with $\sigma^{2(1)}_{1,0,0}=\sigma^{2(2)}_{1,0,0}=\sigma^{2(1)}_{1,1,0}=\sigma^{2(2)}_{1,1,0}=\sigma^{2(1)}_{0,0,0}=\sigma^{2(2)}_{0,0,0}=1$. 
    \item {\bf The case of a binary $Y$.} The parameters are as follows:
    given $W=1$, $\mu_{1,0,1}=0.9$, $\mu_{1,1,1}=0.6$, $\mu_{0,0,1}=0.3$; given $W=0$, $\mu_{1,0,0}=0.2$, $\mu_{1,1,0}=0.1$, $\mu_{0,0,0}=0.8$.
    \item {\bf The case of a Poisson $Y$.} The parameters are as follows:
    given $W=1$, $\mu_{1,0,1}=2$, $\mu_{1,1,1}=1$, $\mu_{0,0,1}=7$; given $W=0$, $\mu_{1,0,0}=5$, $\mu_{1,1,0}=4$, $\mu_{0,0,0}=1$.
    
\end{itemize}

\begin{table}[htb]
\renewcommand{\arraystretch}{0.8}
 \setlength{\belowcaptionskip}{0.2cm}     
 \centering\caption{Bounds on the SACE/$\mbox{SACE}_w$ with Assumptions 3 and 4.}
 \label{table1-4}
 \begin{tabular}{llcc}
    \hline
    Distribution & \quad  & True value  &  Bounds    \\
    \hline
    Normal & $\mbox{SACE}_w$(with cov.) & -0.4 & $[-0.846,-0.050]$\\
    \quad  & $\mbox{SACE}_{w=1}$ & -4 & $[-4.440,-3.644]$ \\
    \quad  & $\mbox{SACE}_{w=0}$ & 2 & $[1.553,2.349]$ \\
    \quad  & $\mbox{SACE}_{\bar{w}}$(without cov.) & -0.4 & $[-1.018,0.377]$\\    
    \hline
    Normal Mixture  & $\mbox{SACE}_w$(with cov.) & 0.56 & $[0.125, 1.337]$\\
    \quad  & $\mbox{SACE}_{w=1}$ & 3.8 & $[3.433,4.394]$ \\
    \quad  & $\mbox{SACE}_{w=0}$ & -1.6 & $[-2.222,-0.867]$ \\
    \quad  & $\mbox{SACE}_{\bar{w}}$(without cov.)  & 0.56 & $[0.0405, 1.583]$\\    
    \hline
    Binary & $\mbox{SACE}_w$(with cov.) & -0.12 & $[-0.200,-0.032]$\\
    \quad   & $\mbox{SACE}_{w=1}$ & 0.6 & $[0.467,0.700]$ \\ 
    \quad   & $\mbox{SACE}_{w=0}$ &-0.6 & $[-0.644,-0.520]$ \\
    \quad   & $\mbox{SACE}_{\bar{w}}$(without cov.) & -0.12 & $[-0.231,0.064]$\\      
    \hline   
    Poisson & $\mbox{SACE}_w$(with cov.) & -1.4 & $[-1.831,-0.687]$\\
    \quad   & $\mbox{SACE}_{w=1}$ & 4 & $[3.560,5.067]$ \\ 
    \quad   & $\mbox{SACE}_{w=0}$ & -5 & $[-5.437,-4.544]$ \\
    \quad   & $\mbox{SACE}_{\bar{w}}$(without cov.) & -1.4 & $[-2.010,-0.513]$\\      
    \hline    
\end{tabular}
\end{table}

The true values and the averages of the estimated upper/lower bounds with a binary covariate are summarized in Table~\ref{table1-4}. In addition, $\mbox{SACE}_{\bar{w}}$ denotes the $\mbox{SACE}$ when the covariate $W$ is ignored. The bounds obtained in 1000 simulation runs 100\% cover the true values in the above four cases. It is also shown that our method not only can correctly identify the sign of the $\mbox{SACE}$ but also can obtain much tighter bounds when the additional information of the covariate $W$ is considered; that is, the bounds on the $\mbox{SACE}$ are a subset of those on $\mbox{SACE}_{\bar{w}}$. For the normal case, it can be seen that the bound intervals for $\mbox{SACE}_{\bar{w}}$ contain zero when the covariate information is not taken into account. This results in the most critical information being lost in practical applications, as it cannot be determined whether the treatment or policy has a positive or negative effect.
\section{Application to NSW data}\label{sec:real-data}
To illustrate our results, we study the causal effect of job training on earnings based on real data from the National Supported Work Demonstration (NSW). The NSW was a temporary experimental program intended to promote reemployment among unemployed workers by providing work experience, which was conducted in the mid-1970s and described by \citet{1986LaLonde}. 
The dataset contains observation samples representing 445 individuals, of whom 185 were randomly assigned to the treatment group (received job training, denoted by $Z=1$) and 260 were randomly assigned to the control group (not received job training, denoted by $Z=0$). To assess the effect of this employment training program on earnings, we take the logarithm of an individual’s earnings in 1978 as the outcome variable, denoted by $Y$, and treat the outcome as analogous to `death' ($S=0$) if the individual was unemployed in 1978. The model is assumed to involve a series of discrete and continuous covariates.  We note that the SACE measures an additive causal effect on the logarithm of earnings, which is equivalent to a multiplicative effect on earnings.

Table~\ref{LL_LD} presents the results for the `always-survivor' and `protected' strata under various assumptions. When only Assumptions 1 and 2 are invoked (without Assumptions 3 and 4), the bounds for the `protected' stratum are wider than those for the `always-survivor' stratum, which might be attributable to the fact that (\ref{fracPf1}) has fewer constraints than (\ref{fracPf}). The effect of job training on individual earnings in the `always-survivor' stratum ranges from a factor of 0.276 to a factor of 4.096. Not surprisingly, Assumption 3 (monotonicity) narrows the bounds substantially for the `always-survivor' stratum, although the identification region still includes zero. As discussed before, Assumption 3 (monotonicity) has no identifying power for the `protected' stratum, as $\pi_{2}=0$ implies the widest bounds possible. Assumption 4 (stochastic dominance) causes the bounds for both strata to be narrower than in the worst case. However, using both assumptions jointly brings important improvements, for the `always-survivor' stratum, the bounds clearly show that the wage effect of job training is positive and ranges from a factor of 1.083 to a factor of 1.501.

\begin{table}[htb]
\renewcommand{\arraystretch}{0.8}
 \setlength{\belowcaptionskip}{0.2cm}     
 \centering\caption{Bounds on the causal effects of job training on earnings (measured on a natural log scale) in the `always-survivor' and `protected' strata without any covariates.}
\label{LL_LD}
\begin{tabular}{lcc}
          \hline
          Assumption & Always-survivor & Protected  \\
          \hline
          None & $[-1.289,1.410]$ & $[-4.262, 4.773]$\\
          Monotonicity & $[-0.166,0.406]$ & 
          $[-4.262, 4.773]$ \\
          Stochastic Dominance & $[-0.516, 0.875]$ &
          $[-4.262, 3.332]$ \\
          Both & 
          $[0.080, 0.406]$ & 
          $[-4.262, 3.332]$  \\
          \hline
\end{tabular}
\end{table}

An important and attractive contribution of our proposed approach, compared to the method of large sample bounds \citep{2014Freiman}, is that it allows the consideration of arbitrary types and numbers of covariates. In general, the use of covariates may help to understand the data in a more detailed way, thus making it possible to obtain narrower bounds compared to the case in which no covariates are considered, and these bounds are also more reliable and valuable. Here, we consider a mix of continuous and binary covariates, including whether the individual is Black (denoted by `Black'), whether the individual is Hispanics (denoted by `Hisp'), whether the individual has a high school degree (denoted by `Nodegree'), the individual's employment status in 1975 (`Unem75'), the individual's marital status (`Married'), and the age and education level of the individual. The results for the bounds are shown in Table~\ref{bounds_multicov}. It can be seen that for the `always-survivor' stratum, the bounds in each case are significantly narrower than the corresponding bounds shown in Table~\ref{LL_LD} when valid covariates are considered. To be specific, the wage effect of job training for the `always-survivor' stratum ranges from a factor of 1.076 to a factor of 1.409 under Assumptions 1-4 rather than from a factor of 1.083 to a factor of 1.501. The more surprising finding is that we can obtain bounds for the `protected' and `harmed' strata at the same time, although the intervals for the `protected' stratum are only slightly tightened.

\begin{table}[htb]
\renewcommand{\arraystretch}{0.8}
 \setlength{\belowcaptionskip}{0.2cm}     
 \centering\caption{Bounds on the ACE in the `always-survivor' and other strata with covariates.
} 
\label{bounds_multicov}
\begin{tabular}{lccc}
          \hline
          Assumption & Always-survivor & Protected & Harmed  \\
          \hline
          None & $[-1.220, 1.273]$ & 
          $[-3.950, 4.288]$ & $[-5.222, 5.417]$\\
          Monotonicity & $[-0.152,0.343]$ & 
          $[-3.950, 4.288]$ & $[-5.222, 5.417]$ \\
          Stochastic Dominance & $[-0.491, 0.764]$ &
          $[-3.950, 2.954]$ & $[-3.052, 5.417]$ \\
          Both & $[0.073, 0.343]$ & 
          $[-3.950, 2.954]$ & $[-3.052, 5.417]$ \\
          \hline
\end{tabular}
\end{table}



\section{Discussion}
In this paper, we explore the bounds on causal effects in scenarios where outcomes are truncated by death under the principal stratification framework introduced by \cite{2002Frangakis}. Most previous research on partial identification has predominantly focused on the `always-survivor' stratum within the principal stratification framework. We introduce a new unified nonparametric framework that provides bounds on the causal effects on both discrete and continuous outcomes within different strata. Our method establishes sharp bounds for not only the SACE (survivor average causal effect) but also other causal effects in other strata. Two fundamental assumptions, namely, the SUTVA and weak ignorability, serve as the primary foundation for our method. Furthermore, additional assumptions such as monotonicity and stochastic dominance are incorporated to broaden the applicability of our bounds. The core concept of our approach is the enhancement of the optimization problem through the utilization of a comprehensive tail probability expectation formula for a set of conditional probability distributions. Challenging situations arise when solving symbolic Balke–Pearl linear programming problems. 
To obtain bounds on the SACE, it is necessary to ensure that $\pi_{0\cdot w} > \pi_{3\cdot w}$, thereby preventing the denominator of the objective function from being zero. Similarly, the ACE bound for the `protected' stratum is informative only when $\pi_{1\cdot w}>\pi_{2\cdot w}$ is satisfied. Otherwise, the bounds on the ACE in the `harmed' stratum become informative. In fact, it is possible to determine in which stratum the bounds on the ACE are valid based on the observed data. For example, if $P_{1\cdot 0w} > P_{0\cdot 1w}$ and $P_{1 \cdot 1w} > P_{1\cdot 0w}$, then our method can be used to obtain bounds on the SACE and $\Delta_1$. The more surprising finding is that when there are continuous covariates, we can obtain bounds for the `protected' and `harmed' strata at the same time. Therefore, a more efficient way of deriving the proportions of each stratum may improve our proposed bounds, and future work can attempt to address this problem.

As illustrated in \ref{Appendix: extension}, this unified framework is not only applicable to situations where outcomes are truncated by death, but can also be extended to data subject to noncompliance. The ACE bounds for different strata are considered in this paper for the case in which the outcome variable $Y$ is a squared-integrable random variable, either discrete or continuous. The covariate types and number are not restricted, enabling the consideration of both binary and continuous scenarios; this allows the ignorability assumption to be relaxed, making the bounds applicable to more situations.

\bibliographystyle{elsarticle-harv}
\bibliography{ref}
\addtolength{\textheight}{-.3in}%

\newpage
\bigskip
\begin{appendices}
\setcounter{table}{0}
\setcounter{figure}{0}
\setcounter{section}{0}
\setcounter{subsection}{0}
\setcounter{equation}{0}
\setcounter{theorem}{0}
\setcounter{page}{1}

\renewcommand{\thesection}{Appendix \Alph{section}}
\renewcommand{\thesubsection}{\Alph{section}.\arabic{subsection}}
\renewcommand{\thetable}{\Alph{section}\arabic{table}}
\renewcommand{\thefigure}{\Alph{section}\arabic{figure}}
\renewcommand{\theequation}{\Alph{section}.\arabic{equation}}
\renewcommand{\thetheorem}{\Alph{section}\arabic{theorem}}
\renewcommand{\thelem}{\Alph{section}\arabic{lem}}

\begin{center}
{\large\bf Supplementary Material for \\
`A unified framework for bounding causal effects on the always survivor and other populations'}
\end{center}

\ref{Appendix: proof} gives all the proofs, \ref{Appendix: extension} shows how the proposed procedure can be extended to the ACE and CACE cases, and \ref{Appendix: simu} supplements the simulation study of more situations, demonstrating the sensitivity of the method proposed in this article to the violation of assumptions 3 and 4 respectively, and finally \ref{Appendix: real data} further analyzes the real dataset from NSW.

\section{Proofs of theoretical results}\label{Appendix: proof}
\subsection{Proof of Proposition 1}
\label{Appendix: pi_k}
As mentioned in section 2, the proportion of the principal stratum is not point-identifiable without the monotonicity assumption. Before discussing the bounds of the causal parameters of interest (such as SACE), the range of $\pi_{k \cdot w}$ need to be derived. For example, consider the problem of finding the bounds on the `harmed' proportion $\pi_{2 \cdot w}$, which is equivalent to solving the following optimization problem:
\begin{equation}
\min\,\mbox{or}\, \max \pi_{2 \cdot w},
\end{equation}
subject to
\begin{eqnarray*}
\left\{
		\begin{aligned}
&              \pi_{0\cdot w}+\pi_{1\cdot w}=P_{1 \cdot 1w},\\
&              \pi_{0\cdot w}+\pi_{2\cdot w}=P_{1 \cdot 0w},\\
&              \pi_{2\cdot w}+\pi_{3\cdot w}=P_{0 \cdot 1w},\\
&              \pi_{1\cdot w}+\pi_{3\cdot w}=P_{0 \cdot 0w},\\
&              \pi_{0\cdot w}+\pi_{1\cdot w}+\pi_{2\cdot w}+\pi_{3\cdot w}=1,\\
&              \pi_{i\cdot w}\geq 0 (i=0,1,2,3).
		\end{aligned}
\right.
\end{eqnarray*}    
After some calculations, we find that
\begin{equation}
   \max\{0,P_{1\cdot 0w}-P_{1\cdot 1w}\}\leq \pi_{2 \cdot w} \leq \min\{P_{1\cdot 0w}, P_{0\cdot 1w}\}.
\end{equation}
Similarly, the following inequalities represent probability bounds in other principal strata:
\begin{eqnarray}
\begin{aligned}
  \max\{0,P_{1\cdot 1w}-P_{0\cdot 0w}\}\leq \pi_{0\cdot w} \leq \min\{P_{1\cdot 0w}, P_{1\cdot 1w}\},\\
  \max\{0,P_{1\cdot 1w}-P_{1\cdot 0w}\}\leq \pi_{1\cdot w} \leq \min\{P_{0\cdot 0w}, P_{1\cdot 1w}\}.
\end{aligned}
\end{eqnarray}

\subsection{Derivation of Equation (\ref{linear programming 1})}
\label{Appendix: equ.(3.4)}
In order to obtain sharp boundaries of causal parameters, the core idea of this article is to construct a Bakle-Pearl linear programming problem. Taking the first equation in constraint (\ref{linear programming 1}) as an example, we have 
\begin{align*}
F_{11 \cdot 0w}&=P (Y \leq y, S=1 \mid Z=0, w) \\
&=P (Y(0) \leq y, S(0)=1 \mid Z=0, w) \\
&=P (Y(0) \leq y, S(0)=1, S(1)=1 \mid  w) + P (Y(0) \leq y, S(0)=1, S(1)=0 \mid  w)\\
&=F_{00\cdot w}+F_{20\cdot w}+(F_{02\cdot w}+F_{22\cdot w})\\
&=F_{00\cdot w}+F_{20\cdot w}+F_{1\cdot 2w}.
\end{align*}
The second equality follows from consistency. The third equality follows because of Assumption~\ref{assu2}. Similarly, we can obtain the following three constraints. The other two constraints are obtained directly from the properties of probability.

\subsection{Balke-Pearl LP method and some Lemmas}
\label{Appendix: lemma}
Here, we first briefly introduce the general situation of how to use the Balke-Pearl linear programming method to solve linear programming problems.

Suppose we have a set of constraints on $\mathbf{x}$  as well as the objective function $c^{\top}\mathbf{x}$ of interest in terms of $\mathbf{x}\in \mathbb{R}^n$:
\begin{eqnarray*}
\min  \quad c^{\top}\mathbf{x}\\\vspace{1cm}
\vspace{1cm}
\text{s.t.} \quad \left\{
  \begin{aligned}
  \sum \mathbf{x} &= 1,\\
      R\mathbf{x} &= p,\\
      A_{l}\mathbf{x} &\leq b_{l},\\
      \mathbf{x}  &\geq 0,
\end{aligned}
\right.
\end{eqnarray*}
where $c\in \{-1,0,1\}^n$, $p\in \mathbb{R}^{m_e}$ and $b_{l}\in \mathbb{R}^{m_l}$ are constants vectors; $R\in \{0,1\}^{m_e\times n}$ and $A_l\in \{0,1\}^{m_l\times n}$ are coefficient matrix. By the strong duality theorem of convex optimization, the optimal value of this primal problem is equal to that of its dual. To simplify the notation, we let $A_e:=\left(1_{n \times 1} \quad R^\top \right)^\top$, and 
$b_e:=\left(1 \quad p\right)^\top$; then, we have
\begin{small}
\begin{align*}
& \min \left\{c^\top \mathbf{x} \mid \mathbf{x} \in \mathbb{R}^n, A_{l} \mathbf{x} \leq b_{l}, A_e \mathbf{x}=b_e, \mathbf{x} \geq 0_{n \times 1}\right\} \\
 =&\max \left\{\left(\begin{array}{cc}
b_{l}^\top & b_e^\top
\end{array}\right) \mathbf{y} \mid  \mathbf{y} \in \mathbb{R}^{m_l+m_e},\left(\begin{array}{cc}
A_{l}^\top & A_e^\top
\end{array}\right)  \mathbf{y} \leq c,\left(I_{m_l \times m_l} \quad 0_{m_l \times m_e}\right)  \mathbf{y} \leq 0_{m_l \times 1}\right\}\\
=&\max \left\{\left(\begin{array}{ll}
b_{l}^T & b_e^T
\end{array}\right) \bar{ \mathbf{y}} \mid \bar{ \mathbf{y}} \text { is a vertex of }\left\{ \mathbf{y} \in \mathbb{R}^{m_l+m_e} {\bigg|}\left(\begin{array}{cc}
A_{l}^\top & A_e^\top \\
I_{m_l \times m_l} & 0_{m_l \times m_e}
\end{array}\right)  \mathbf{y} \leq\left(\begin{array}{c}
c \\
0_{m_l \times 1}
\end{array}\right)\right\}\right\}
\end{align*}
\end{small}
To solve this symbolic Balke–Pearl linear programming problem, we use the R package `rcdd' based on enumerating the vertices of the constraint polygon of the dual problem. Whereas some programs that can only provide numerical solutions for specific data, this R package takes a symbolic description as input, and its output is a symbolic solution.

As we mentioned before, we cannot use the linear programming method of \citet{1997Balke} directly,
since conditioning on being selected into the study generally implies a nonlinear structure on the counterfactual probabilities. Thus, before proving Theorem~\ref{thm-SACE}, we provide some lemmas to adapt the symbolic Balke-Pearl linear programming method to situations where the objective function is nonlinear, and show the connection between the fractional programming problem and a certain parametric linear programming problem.

\begin{lem}\label{lemma: frac}
\citep{1983Schaible}
{ The linear fractional programming problem}
\[\sup\Big\{ \frac{\boldsymbol\gamma^\top\mathbf{x}+a}{\boldsymbol\alpha ^{\top}\mathbf{x}+b}: \mathbf{Ax} \leq \boldsymbol\beta, \boldsymbol\alpha ^{\top}\mathbf{x} +b > 0, \mathbf{x} \geq \mathbf{0}\Big\}\]
\emph{is equivalent to the linear programming problem.}
\[\sup\Big\{\boldsymbol\gamma^\top \mathbf{y}+at: \mathbf{Ay}-\boldsymbol\beta t \leq \mathbf{0}, \boldsymbol\alpha ^{\top} \mathbf{y}+bt = \mathbf{1}, \mathbf{y} \geq \mathbf{0}, t >0 \Big\}\]
\emph{where $\mathbf{y}=\mathbf{x}/(\boldsymbol\alpha^\top\mathbf{x}+b)$, $t=1/(\boldsymbol\alpha^\top \mathbf{x}+b)$; $\mathbf{A}$ is a coefficient matrix; $\boldsymbol\alpha$, $\boldsymbol\beta$, and  $\boldsymbol\gamma$ are constants vectors; $\mathbf{x}$ is a vector of variables; $a$ and $b$ are constants; and $\mathbf{x} \geq \mathbf{0}$ indicates that each component of $\mathbf{x}$ is greater than or equal to 0.}
\end{lem} 

In the specified case, for example, programming problem (\ref{fracP}), the corresponding setting coincides with the above theorem as follow: 
\[\mathbf{x} = \left(F_{00\cdot w},F_{10\cdot w},F_{20\cdot w},F_{30.w},F_{0.1w},F_{3\cdot 1w},F_{1 \cdot 2w},F_{2\cdot2w},F_{\cdot3w}\right)^\top,\quad \mathbf{y}=\mathbf{x}t^{(0)}\]
\[a=b=0,\quad \boldsymbol\gamma=(0,-1,1,0,0,0,0,0,0)^\top,\quad \boldsymbol\alpha=(1,1,1,1,0,0,0,0,0)^\top\]
\[\mathbf{A}= \begin{bmatrix}
                1 & 0 & 1 & 0 & 0 & 0 & 1 & 0 & 0\\
                0 & 1 & 0 & 1 & 0 & 0 & 0 & 1 & 0\\
                1 & 1 & 0 & 0 & 1 & 0 & 0 & 0 & 0\\
                0 & 0 & 1 & 1 & 0 & 1 & 0 & 0 & 0\\
                1 & 1 & 1 & 1 & 1 & 1 & 1 & 1 & 1
            \end{bmatrix}, \quad 
  \boldsymbol\beta = \left(F_{11\cdot 0w},F_{01\cdot0w},F_{11\cdot1w},F_{01\cdot1w},1\right)^\top\]
and $t^{(0)}=1/(F_{00 \cdot w} + F_{10 \cdot w} + F_{20 \cdot w} +F_{30 \cdot w})$. Let $f_{ij\cdot w}^{(0)}=F_{ij\cdot w}/(F_{00 \cdot w} + F_{10 \cdot w} + F_{20 \cdot w} +F_{30 \cdot w})= F_{ij\cdot w}\cdot t^{(0)}$; similarly, $f_{0\cdot 1\cdot w}=F_{0\cdot 1\cdot w}\times t^{(0)}$, $f_{3\cdot 1w}^{(0)}=F_{3\cdot 1w}\times  t^{(0)}$, $f_{1\cdot 2w}^{(0)}=F_{1\cdot 2w}\times  t^{(0)}$, $f_{2\cdot 2w}^{(0)}=F_{2\cdot 2w}\times  t^{(0)}$ and $f_{\cdot 3w}^{(0)}=F_{\cdot 3w}\times  t^{(0)}$.  

In addition, \citet{1962Dinkelbach} reduces the solution of a linear fractional programming problem to the solution of a sequence of linear programming problems. To be specific, consider a typical linear fractional programming problem:
\begin{equation}
\label{fractional programming}
    \max\Big\{ \lambda = \frac{f(\mathbf{x})}{g(\mathbf{x})}: A\mathbf{x} = b, x \geq 0\Big\}
\end{equation}
Let us assume that the domain of feasible solutions, $D$, is nonempty; the denominator does not reduce to a constant; and that it is strictly positive on $D$. Consider the auxiliary function
\[F(\lambda) = \max_{\mathbf{x} \in D} [f(\mathbf{x}) - \lambda g(\mathbf{x})], \quad\quad \lambda \in R.\]

\begin{lem}\label{lemma: equivalence}
{ The vector $\mathbf{x}_0$ is the optimal solution to  the fractional programming problem given in (\ref{fractional programming}) if and only if} 
\[ \max_{\mathbf{x} \in D} [f(\mathbf{x}) - \lambda_0 g(\mathbf{x})] = F(\lambda_0) = 0 ,\]
where
\[ \lambda_0 = \frac{f(\mathbf{x}_0)}{g(\mathbf{x}_0)}\]
\end{lem}
For solving the symbolic linear programming problem in (\ref{fracP}), this corresponds to taking $f(\mathbf{x})=F_{20\cdot w} - F_{10\cdot w}$, $g(\mathbf{x})=F_{00\cdot w}+F_{10\cdot w}+F_{20\cdot w}+F_{30\cdot w}$ and $\lambda_0$ denotes the upper bound obtained by fixing $y_0$. The lower bound in (\ref{fracP}) can also be assessed analogously.

\subsection{Proof of Theorem~\ref{thm-SACE}}
\label{Appendix: thm-SACE}
\begin{proof}
As shown in the main text, we can obtain the sharp bounds of SACE under Assumption~\ref{assu1} and \ref{assu2} by minimizing or maximizing (\ref{fracP}) subject to (\ref{linear programming 1}).
Note that $\pi_{0\cdot w} =  F_{11\cdot0w}+F_{01\cdot0w}-\pi_{2\cdot w}=P_{1\cdot 0w}-\pi_{2\cdot w}$. By using Lemma~\ref{lemma: frac}, the linear fractional programming problem in (\ref{fracP}) can be transformed into a linear programming problem:
\begin{equation}\label{fracPf}
\min\,\mbox{or}\, \max f_{20\cdot w}^{(0)}-f_{10\cdot w}^{(0)},
\end{equation}
subject to
\begin{eqnarray*}
\left\{
		\begin{aligned}
		&f_{00 \cdot w}^{(0)}+f_{20 \cdot w}^{(0)}+f_{1 \cdot 2w}^{(0)} =  {F_{11 \cdot 0w}}/(P_{1\cdot 0w}-\pi_{2\cdot w}), \\
              &f_{10 \cdot w}^{(0)}+f_{30 \cdot w}^{(0)}+f_{2 \cdot 2w}^{(0)} =   {F_{01 \cdot 0w}}/(P_{1\cdot 0w}-\pi_{2\cdot w}), \\
              &f_{00 \cdot w}^{(0)}+f_{10 \cdot w}^{(0)}+f_{0 \cdot 1w}^{(0)} =   {F_{11 \cdot 1w}}/(P_{1\cdot 0w}-\pi_{2\cdot w}),\\
             & f_{20 \cdot w}^{(0)}+f_{30 \cdot w}^{(0)}+f_{3 \cdot 1w}^{(0)} =   {F_{01 \cdot 1w}}/(P_{1\cdot 0w}-\pi_{2\cdot w}),\\
             & f_{00 \cdot w}^{(0)}+f_{10 \cdot w}^{(0)}+f_{20 \cdot w}^{(0)}+f_{30 \cdot w}^{(0)}=1 ,\\
             & \sum_{i=0}^3 f_{i0 \cdot w}^{(0)}+f_{0 \cdot 1w}^{(0)}+f_{3 \cdot 1w}^{(0)}+f_{1\cdot 2w}^{(0)}+f_{2\cdot 2w}+f_{\cdot 3w}^{(0)} =   {1}/(P_{1\cdot 0w}-\pi_{2\cdot w}), \\
            &  f_{i0 \cdot w}^{(0)} \geq 0 (i=0,1,2,3), f_{0\cdot 1w}^{(0)}\geq 0, f_{3\cdot 1w}^{(0)}\geq 0,f_{1\cdot 2w}^{(0)}\geq 0, f_{2\cdot 2w}^{(0)}\geq 0, f_{\cdot 3w}^{(0)}\geq 0,
        \end{aligned}
\right.
\end{eqnarray*}
where $t^{(0)}=1/(F_{00 \cdot w}+ F_{10 \cdot w}+ F_{20 \cdot w}+F_{30 \cdot w}),f_{ij\cdot w}^{(0)}=t^{(0)}F_{ij\cdot w}$, similarly, $f_{0\cdot 1w}^{(0)}=t^{(0)}F_{0\cdot 1w}$, $f_{3\cdot 1w}^{(0)}=t^{(0)}F_{3\cdot 1w}$, $f_{1\cdot 2w}^{(0)}=t^{(0)}F_{1\cdot 2w}$, $f_{2\cdot 2w}^{(0)}=t^{(0)}F_{2\cdot 2w}$ and  $f_{\cdot 3w}^{(0)}=t^{(0)}F_{\cdot 3w}$. 
Given $\pi_{2\cdot w}$, we use the symbolic Balke–Pearl linear programming method to solve the linear programming problem in (\ref{fracPf}), which yields a closed-form solution. 

As shown in the general case in the Appendix~\ref{Appendix: lemma}, according to the strong duality theorem of convex optimization, the optimal value of this primal problem is equal to that of its dual \citep{2007LP}. Furthermore, its constraint space is a convex polygon, and by the fundamental theorem of linear programming, this optimum is attained at one of its vertices. By plugging the vertices into the dual objective function and evaluating the expression, we obtain

\begin{equation}\label{bound_f0}
L_0(y,w|\pi_{2\cdot w})\leq f_{20\cdot w}^{(0)}-f_{10\cdot w}^{(0)}\leq R_0(y,w|\pi_{2\cdot w}), 
\end{equation}
where
\begin{eqnarray*}
R_0(y,w|\pi_{2\cdot w})= \min \left\{
\frac {F_{01 \cdot 1w}}{P_{1\cdot 0 w}-\pi_{2\cdot w}},
 \frac {F_{11 \cdot 0w}}{P_{1\cdot 0 w}-\pi_{2\cdot w}},
 \frac {F_{01 \cdot 1w}+F_{11 \cdot 0w}}{P_{1\cdot 0 w}-\pi_{2\cdot w}} - 1,
 1
\right\},
\end{eqnarray*}
and
\begin{eqnarray*}
L_0(y,w|\pi_{2\cdot w})= \max \left\{
     \frac {-F_{11 \cdot 1w}}{P_{1\cdot 0 w}-\pi_{2\cdot w}}, 
\frac {-F_{01 \cdot 0w}}{P_{1\cdot 0 w}-\pi_{2\cdot w}},
 \frac {-(F_{01 \cdot 0w}+F_{11 \cdot 1w})}{P_{1\cdot 0 w}-\pi_{2\cdot w}} + 1,
       -1
\right\}.
\end{eqnarray*}
When $\pi_{2\cdot w}$ is varying on its support, we have
\begin{equation*}
\inf_{\pi_{2\cdot w}} L_0(y,w|\pi_{2\cdot w}) \leq f_{20\cdot w}^{(0)}-f_{10\cdot w}^{(0)} \leq  \sup_{\pi_{2\cdot w}} R_0(y,w|\pi_{2\cdot w}). 
\end{equation*} 
At the same time, we know that the bounds on $\mbox{SACE}_w$ defined in (\ref{SACEw}) can be written as
\begin{equation}\label{b1_SACEw}
\int_{-\infty}^\infty L_0(y,w|\pi_{2\cdot w}) dy\leq \mbox{SACE}_w\leq \int_{-\infty}^\infty R_0(y,w|\pi_{2\cdot w})dy.
\end{equation}
Note that the term $R_0(y,w|\pi_{2\cdot w})$ is increasing with $\pi_{2\cdot w}$ and that $L_0(y,w|\pi_{2\cdot w})$ is decreasing with $\pi_{2\cdot w}$. Moreover, recall the bounds on $\pi_{2\cdot w}$ given in Proposition \ref{prop1}. Accordingly, we conclude from (\ref{b1_SACEw}) as shown in (\ref{bound_00}) with $\pi_{2\cdot w}^{\scriptstyle\max}=\min\{P_{1\cdot 0w}, P_{0\cdot 1w}\},$ which is the maximum value of $\pi_{2\cdot w}.$ By the law of total expectation, we have $\mbox{SACE} ={E\{\pi_{0 \cdot W}\mbox{SACE}_W\}}/{E(\pi_{0 \cdot W})}.$
Then, from (\ref{bound_00}) we obtain the following:
\begin{equation*}
   \frac {E\Big\{\pi_{0 \cdot W}\int_{-\infty}^{+\infty}L_0(y,W|\pi_{2\cdot W})dy\Big\}}{E\pi_{0 \cdot W}}\leq \mbox{SACE} \leq \frac {E\Big\{\pi_{0 \cdot W}\int_{-\infty}^{+\infty}R_0(y,W|\pi_{2\cdot W})dy\Big\}}{E\pi_{0 \cdot W}}.
\end{equation*}
Since the above right-hand and left-hand inequalities take the equal sign when $\pi_{2\cdot W}=\pi_{2\cdot W}^{\scriptstyle\max},$ that is, $\pi_{0\cdot W}=P_{1\cdot 0 W}-\pi_{2\cdot W}^{\scriptstyle\max},$ the above inequality reduces to 
\begin{equation}
 L_0\leq \mbox{SACE} \leq R_0,
\end{equation}
with 
\begin{eqnarray*}
    \begin{aligned}
        &L_0=\frac {E\Big\{(P_{1\cdot 0 W}-\pi_{2 \cdot W}^{\scriptstyle\max})\int_{-\infty}^{+\infty}L_0(y,W|\pi_{2\cdot W}^{\scriptstyle\max})dy\Big\}}{E(P_{1\cdot 0 W}-\pi_{2 \cdot W}^{\scriptstyle\max})},\\
        &R_0=\frac {E\Big\{(P_{1\cdot 0 W}-\pi_{2 \cdot W}^{\scriptstyle\max})\int_{-\infty}^{+\infty}R_0(y,W|\pi_{2\cdot W}^{\scriptstyle\max})dy\Big\}}{E(P_{1\cdot 0 W}-\pi_{2 \cdot W}^{\scriptstyle\max})}.
    \end{aligned}
\end{eqnarray*}
\end{proof}

\subsection{Proof of Theorem~\ref{thm-Delta12}}
\label{Appendix: thm-Delta12}
\begin{proof}
We first consider $\Delta_{1\cdot w}$.  Let us define $t^{(1)} = 1/(F_{01\cdot w}+F_{11\cdot w}+F_{21\cdot w}+F_{31\cdot w}), f_{ij\cdot w}^{(1)}= t^{(1)}F_{ij\cdot w}$, $f_{1.2w}^{(1)} = f_{02.w}^{(1)}+f_{22.w}^{(1)}$, $f_{2.2w}^{(1)} = f_{12.w}^{(1)}+f_{32.w}^{(1)}$ and $f_{.3w}^{(1)}=\sum_{i=0}^3 f_{i3.w}^{(1)}$. Recall that $F_{a1\cdot zw}=\P(I(Y\leq y)=a,S=1 \mid Z=z,W=w)$,  $a,z=0,1,$ and let $F_{a0\cdot zw}$ denote the observed conditional joint probability distribution $\P(I(Y\leq y)=a,S=0 \mid Z=z,W=w)$, $a,z=0,1.$ The optimization of (\ref{linear_programming_delta1}) can be transformed into the following symbolic linear programming problem:
\begin{equation}\label{fracPf1}
\min\,\mbox{or}\, \max f_{21\cdot w}^{(1)}-f_{11\cdot w}^{(1)},
\end{equation}
subject to
\begin{eqnarray*}
\label{LP_LD_missing}
\left\{
            \begin{aligned}
            & f_{00 \cdot w}^{(1)}+f_{20 \cdot w}^{(1)}+f_{1 \cdot 2w}^{(1)} = {F_{11 \cdot 0w}}/\left({P_{1\cdot 1w}-P_{1\cdot 0w} + \pi_{2\cdot w}}\right),\\
            & f_{10 \cdot w}^{(1)}+f_{30 \cdot w}^{(1)}+f_{2 \cdot 2w}^{(1)} = {F_{01 \cdot 0w}}/\left({P_{1\cdot 1w}-P_{1\cdot 0w} + \pi_{2\cdot w}}\right),\\
             & f_{01 \cdot w}^{(1)}+f_{11 \cdot w}^{(1)}+f_{00 \cdot w}^{(1)}+f_{10 \cdot w}^{(1)}= {F_{11 \cdot 1w}}/\left({P_{1\cdot 1w}-P_{1\cdot 0w} + \pi_{2\cdot w}}\right),\\ 
             & f_{21 \cdot w}^{(1)}+f_{31 \cdot w}^{(1)} +f_{20 \cdot w}^{(1)}+f_{30 \cdot w}^{(1)}=  {F_{01 \cdot 1w}}/ \left({P_{1\cdot 1w}-P_{1\cdot 0w} + \pi_{2\cdot w}}\right),\\             
             & f_{01 \cdot w}^{(1)}+f_{11 \cdot w}^{(1)}+f_{21 \cdot w}^{(1)}+f_{31 \cdot w}^{(1)} = 1, \\              
             & \sum_{j=0}^1\sum_{i=0}^3 f_{ij \cdot w}^{(1)}+f_{1 \cdot 2w}^{(1)}+f_{2 \cdot 2w}^{(1)}+f_{\cdot 3w}^{(1)} =  {1}/\left({P_{1\cdot 1w}-P_{1\cdot 0w} + \pi_{2\cdot w}}\right), \\             
             & f_{ij \cdot w}^{(1)} \geq 0\,\,(i=0,1,2,3,j=0,1),f_{1\cdot 2w}^{(1)}\geq 0, f_{2\cdot 2w}^{(1)}\geq 0, f_{\cdot 3w}^{(1)}\geq 0.
            \end{aligned}
\right.
\end{eqnarray*}
Additionally, a symbolic Balke-Pearl linear programming method is used to solve the above linear programming problem, which yields the following solution,
\begin{equation*}
L_1(y,w|\pi_{2\cdot w})\leq f_{21\cdot w}^{(1)}-f_{11\cdot w}^{(1)}\leq R_1(y,w|\pi_{2\cdot w}), 
\end{equation*}
where
\begin{eqnarray*}
R_1(y,w|\pi_{2\cdot w})= \min \left\{
\frac {F_{01 \cdot 1w}}{P_{1\cdot 1w}-P_{1\cdot 0w} + \pi_{2\cdot w}},
1
\right\},
\end{eqnarray*}
and
\begin{eqnarray*}
L_1(y,w|\pi_{2\cdot w})= \max \left\{
\frac {-F_{11 \cdot 1w}}{P_{1\cdot 1w}-P_{1\cdot 0w} + \pi_{2\cdot w}}, 
      -1
\right\}.
\end{eqnarray*}
Note that the obtained right bound is increasing with $\pi_{2\cdot w}$ and that the left bound is decreasing with $\pi_{2\cdot w}$. We have $\inf_{\pi_{2\cdot w}} L_1(y,w|\pi_{2\cdot w})=L_1(y,w|\pi_{2\cdot w}^{\scriptstyle\min})$ and $\sup_{\pi_{2\cdot w}} R_1(y,w|\pi_{2\cdot w})=R_1(y,w|\pi_{2\cdot w}^{\scriptstyle\min})$. Then,
\begin{equation}\label{delta1}
 L_1(y,w|\pi_{2\cdot w}^{\scriptstyle\min})\leq f_{21\cdot w}^{(1)}-f_{11\cdot w}^{(1)} \leq R_1(y,w|\pi_{2\cdot w}^{\scriptstyle\min}).
\end{equation} 
Recall the result given in Proposition \ref{prop1}, $\pi_{2\cdot w}^{\scriptstyle\min}=\max\{0,P_{1\cdot 0w}- P_{1\cdot 1w}\}$, which is the minimum value of $\pi_{2\cdot w}$. We now apply the same argument as for (\ref{bound_00}) and (\ref{bound_00_SACE}) and note that $\Delta_1={E\{\pi_{1 \cdot W}\Delta_{1,W}\}}/{E(\pi_{1 \cdot W})}$ and $\pi_{1 \cdot w}=P_{1\cdot 1w}-P_{1\cdot 0w} + \pi_{2\cdot w}^{\scriptstyle\min}$, thus obtaining the following bounds:
\begin{equation}\label{bound_10_w}
 \int_{-\infty}^{+\infty}  L_1(y,w|\pi_{2\cdot w}^{\scriptstyle\min})\, dy\leq \Delta_{1\cdot w } \leq \int_{-\infty}^{+\infty}  R_1(y,w|\pi_{2\cdot w}^{\scriptstyle\min})\, dy,
\end{equation} 
and
\begin{equation}\label{bound_10}
 L_1\leq \Delta_1 \leq R_1,
\end{equation}
with 
\begin{eqnarray*}
    \begin{aligned}
        &L_1=\frac {E\Big\{(P_{1\cdot 1W}-P_{1\cdot 0W} + \pi_{2\cdot W}^{\scriptstyle\min})\int_{-\infty}^{+\infty}L_1(y,W|\pi_{2\cdot W}^{\scriptstyle\min})dy\Big\}}{E(P_{1\cdot 1W}-P_{1\cdot 0W} + \pi_{2\cdot w}^{\scriptstyle\min})},\\
        &R_1=\frac {E\Big\{(P_{1\cdot 1W}-P_{1\cdot 0W} + \pi_{2\cdot W}^{\scriptstyle\min})\int_{-\infty}^{+\infty}R_1(y,W|\pi_{2\cdot W}^{\scriptstyle\min})dy\Big\}}{E(P_{1\cdot 1W}-P_{1\cdot 0W} + \pi_{2\cdot W}^{\scriptstyle\min})}.
    \end{aligned}
\end{eqnarray*}
Likewise, the conditional ACEs  for the `harmed' strata given by (\ref{delta2}). Let $t^{(2)} = 1/(F_{02\cdot w}+F_{12\cdot w}+F_{22\cdot w}+F_{32\cdot w})$, $f_{ij\cdot w}^{(2)} = t^{(2)} F_{ij\cdot w}$, $f_{0.1w}^{(2)}  = f_{01.w}^{(2)} +f_{11.w}^{(2)}$, $f_{3.1w}^{(2)}  = f_{21.w}^{(2)} +f_{31.w}^{(2)}$ and $f_{.3w}^{(2)} =\sum_{i=0}^3 f_{i3.w}^{(2)} $. Optimizing (\ref{delta2}) leads to the following symbolic linear programming problem:
\begin{equation}\label{fracPf2}
\min\,\mbox{or}\, \max f_{22\cdot w}^{(2)} -f_{12\cdot w}^{(2)} ,
\end{equation}
subject to
\begin{eqnarray*}
\left\{
            \begin{aligned}
             & f_{02 \cdot w}^{(2)} +f_{22 \cdot w}^{(2)}  + f_{00 \cdot w}^{(2)}+f_{20 \cdot w}^{(2)} = F_{11 \cdot 0w}/\pi_{2\cdot w}, \\     
             & f_{12 \cdot w}^{(2)} +f_{32 \cdot w}^{(2)}  + f_{00 \cdot w}^{(2)}+f_{30 \cdot w}^{(2)}  =  F_{01 \cdot 0w}/ \pi_{2\cdot w}, \\
            & f_{00 \cdot w}^{(2)}+f_{10 \cdot w}^{(2)}+f_{0 \cdot 1w}^{(2)}= {F_{11 \cdot 1w}}/\pi_{2\cdot w},\\ 
            & f_{20 \cdot w}^{(2)}+f_{30 \cdot w}^{(2)}+f_{3 \cdot 1w}^{(2)}= {F_{01 \cdot 1w}}/\pi_{2\cdot w},\\ 
             & f_{02 \cdot w}^{(2)} +f_{12 \cdot w}^{(2)} +f_{22 \cdot w}^{(2)} +f_{32 \cdot w}^{(2)} =1, \\ 
             & \sum_{j=0,2}\sum_{i=0}^3 f_{ij \cdot w}^{(2)} +f_{0 \cdot 1w}^{(2)} +f_{3 \cdot 1w}^{(2)} +f_{\cdot 3w}^{(2)}  = 1/\pi_{2\cdot w}, \\
             & f_{ij \cdot w}^{(2)}  \geq 0 \,(i=0,1,2,3,j=0,2),f_{0 \cdot 1w}^{(2)}  \geq 0,f_{3 \cdot 1w}^{(2)}  \geq 0,f_{\cdot 3w}^{(2)} \geq 0.
            \end{aligned}
\right.
\end{eqnarray*}  
Solving the above linear programming problem yields
\begin{equation*}
L_2(y,w|\pi_{2\cdot w})\leq f_{22\cdot w}^{(2)} -f_{12\cdot w}^{(2)} \leq R_2(y,w|\pi_{2\cdot w}), 
\end{equation*}
where
\begin{eqnarray*}
R_2(y,w|\pi_{2\cdot w})= \min \left\{
\frac {F_{11 \cdot 0w}}{ \pi_{2\cdot w}},1 
\right\}\,\,\mbox{and}\,\,L_2(y,w|\pi_{2\cdot w})= \max \left\{
\frac {-F_{01 \cdot 0w}}{ \pi_{2\cdot w}},-1
\right\}.
\end{eqnarray*}
Note the monotonicity of $L_2$ and $R_2$ with respect to $\pi_{2\cdot w}$. Then $\inf_{\pi_{2\cdot w}} L_2(y,w|\pi_{2\cdot w})=L_2(y,w|\pi_{2\cdot w}^{\scriptstyle\min})$ and $\sup_{\pi_{2\cdot w}} R_2(y,w|\pi_{2\cdot w})=R_2(y,w|\pi_{2\cdot w}^{\scriptstyle\min})$. Hence we see that
\begin{equation}\label{bound_DL}
 L_2(y,w|\pi_{2\cdot w}^{\scriptstyle\min})\leq f_{22\cdot w}^{(2)} -f_{12\cdot w}^{(2)}  \leq R_2(y,w|\pi_{2\cdot w}^{\scriptstyle\min}),
\end{equation} 
with $\pi_{2\cdot w}^{\scriptstyle\min}=\max\{0,P_{1\cdot 0w}- P_{1\cdot 1w}\}.$ These bounds are informative only if $P_{1\cdot 0w} > P_{1\cdot 1w}.$ 
In the same manner as for (\ref{bound_00}) and (\ref{bound_00_SACE}), while also noting that $\Delta_2={E\{\pi_{2 \cdot W}\Delta_{2,W}\}}/{E(\pi_{2\cdot W})}$, we obtain the following bounds,
\begin{equation}\label{bound_01_w}
 \int_{-\infty}^{+\infty}  L_2(y,w|\pi_{2\cdot w}^{\scriptstyle\min})\, dy\leq \Delta_{2\cdot w } \leq \int_{-\infty}^{+\infty}  R_2(y,w|\pi_{2\cdot w}^{\scriptstyle\min})\, dy,
\end{equation} 
and
\begin{equation}\label{bound_01}
 \frac {E\Big\{\pi_{2\cdot W}^{\scriptstyle\min}\int_{-\infty}^{+\infty}L_2(y,W|\pi_{2\cdot W}^{\scriptstyle\min})dy\Big\}}{E(\pi_{2\cdot W}^{\scriptstyle\min})}\leq \Delta_2\leq \frac {E\Big\{\pi_{2\cdot W}^{\scriptstyle\min}\int_{-\infty}^{+\infty}R_2(y,W|\pi_{2\cdot W}^{\scriptstyle\min})dy\Big\}}{E(\pi_{2\cdot W}^{\scriptstyle\min})}.
\end{equation}
\end{proof}

\subsection{Proof of Theorem \ref{thm-SACE-m}}
\label{Appendix: thm-SACE-m}
\begin{proof}
Under Assumption~\ref{assu1}-\ref{assump3}, we first let $t_m^{(0)}=1/(F_{11 \cdot 0w}+F_{01 \cdot 0w})$, $m_{ij\cdot w}^{(0)}=t_m^{(0)} F_{ij\cdot w}$, $m_{0\cdot 1w}^{(0)}=t_m^{(0)}F_{0\cdot 1w}$ with $F_{0\cdot 1w}=F_{01\cdot w}+F_{11\cdot w}$, $m_{3\cdot 1w}^{(0)}=t_m^{(0)}F_{3\cdot 1w}$ with $F_{3\cdot 1w}=F_{21\cdot w}+F_{31\cdot w}$, and $m_{\cdot 3w}^{(0)}=t_m^{(0)}F_{\cdot 3w}$ with $F_{\cdot 3w}=F_{03\cdot w}+F_{13\cdot w}+F_{23\cdot w}+F_{33\cdot w}$. Then, the bounding problem for $\mbox{SACE}_w$ is equivalent to the following optimization problem with a more simplified version of the constraints compared to (\ref{fracPf}):
\begin{equation}\label{fracPf0_mon}
\min\,\mbox{or}\, \max m_{20\cdot w}^{(0)}-m_{10\cdot w}^{(0)},
\end{equation}
subject to
\begin{eqnarray*}
\left\{
		\begin{aligned}
		&m_{00 \cdot w}^{(0)}+m_{20 \cdot w}^{(0)} = {F_{11 \cdot 
             0w}}/P_{1\cdot 0w}, \\
            &m_{10 \cdot w}^{(0)}+m_{30 \cdot w}^{(0)} = {F_{01 \cdot 0w}}/P_{1\cdot 0w}, \\
            &m_{00 \cdot w}^{(0)}+m_{10 \cdot w}^{(0)}+m_{0 \cdot 1w}^{(0)} =   {F_{11 \cdot 1w}}/P_{1\cdot 0w},\\
            & m_{20 \cdot w}^{(0)}+m_{30 \cdot w}^{(0)}+m_{3 \cdot 1w}^{(0)} =   {F_{01 \cdot 1w}}/P_{1\cdot 0w},\\
            & m_{00 \cdot w}^{(0)}+m_{10 \cdot w}^{(0)}+m_{20 \cdot w}^{(0)}+m_{30 \cdot w}^{(0)}=1 ,\\
            & \sum_{i=0}^3 m_{i0 \cdot w}^{(0)}+m_{0 \cdot 1w}^{(0)}+m_{3 \cdot 1w}^{(0)}+m_{\cdot 3w}^{(0)} =   {1}/P_{1\cdot 0w}, \\
            &  m_{i0 \cdot w}^{(0)} \geq 0 \,\,(i=0,1,2,3), m_{0\cdot 1w}^{(0)}\geq 0, m_{3\cdot 1w}^{(0)}\geq 0, m_{\cdot 3w}^{(0)}\geq 0.
        \end{aligned}
\right.
\end{eqnarray*}
Unlike (\ref{fracPf}), the above linear programming  problem does not involve any parameters such as $\pi_{k \cdot w}$, $k\in \{0,1,3\}$. Using the method proposed by \citet{1997Balke}, the bounds on the integrand of $\mbox{SACE}_w$ can be obtained as follows:
\begin{equation}\label{SACE_m}
L_{0,m}(y,w)\leq m_{20\cdot w}^{(0)}-m_{10\cdot w}^{(0)}\leq R_{0,m}(y,w), 
\end{equation}
where
\begin{eqnarray*}
\begin{aligned}
& L_{0,m}(y,w)=\max\left\{-\frac {F_{01 \cdot 0w}}{P_{1\cdot 0w}}, \frac {F_{11 \cdot 0w}-F_{11 \cdot 1w}}{P_{1\cdot 0w}}\right\},\\
& R_{0,m}(y,w)= \min\left\{\frac {F_{11 \cdot 0w}}{P_{1\cdot 0w}}, \frac {F_{01 \cdot1w}-F_{01 \cdot 0w}}{P_{1\cdot 0w}}\right\}.
\end{aligned}
\end{eqnarray*}
Note that $\mbox{SACE} ={E\{(F_{11 \cdot 0W}+F_{01 \cdot 0W})\mbox{SACE}_W\}}/{E(F_{11 \cdot 0W}+F_{01 \cdot 0W})}$; thus, we conclude that,
\begin{equation}
 \int_{-\infty}^{+\infty}  L_{0,m}(y,w)\, dy\leq \mbox{SACE}_{w } \leq \int_{-\infty}^{+\infty}  R_{0,m}(y,w)\, dy,
\end{equation} 
and
\begin{equation}
L_{0,m} \leq \mbox{SACE} \leq R_{0,m},
\end{equation}
where
\begin{eqnarray*}
\begin{aligned}
&L_{0,m}=\frac{E\Big\{P_{1\cdot 0W}\int_{-\infty}^{+\infty}  L_{0,m}(y,W)\, dy\Big\}}{E(P_{1\cdot 0W})},\,\,
R_{0,m}=\frac{E\Big\{P_{1\cdot 0W}\int_{-\infty}^{+\infty}  R_{0,m}(y,W)\, dy\Big\}}{E(P_{1\cdot 0W})}.
\end{aligned}
\end{eqnarray*}
\end{proof}

\subsection{Proof of Theorem~\ref{thm-Delta1-m}}
\label{Apppendix:thm-Delta1-m}
\begin{proof}
Under Assumption~\ref{assu1}-\ref{assump3}, we rewrite $\pi_{1\cdot w}=P_{1\cdot 1w}-P_{1\cdot 0w}$. Let $t_m^{(1)} = 1/(F_{01\cdot w}+F_{11\cdot w}+F_{21\cdot w}+F_{31\cdot w})$, $m_{ij\cdot w}^{(1)}= t_m^{(1)}F_{ij\cdot w}$, and $m_{\cdot 3w}^{(1)} = \sum_{i=0}^3m_{i3\cdot w}^{(1)}$. To obtain the bounds on ACE in the ‘protected’ strata, we maximize or minimize $m_{21.w}^{(1)}-m_{11.w}^{(1)}$ subject to
\begin{eqnarray}
\label{fracPf1_mon}
\left\{
            \begin{aligned}
            & m_{00 \cdot w}^{(1)}+m_{20 \cdot w}^{(1)} = {F_{11 \cdot 0w}}/(P_{1\cdot 1w}-P_{1\cdot 0w}),\\
            & m_{10 \cdot w}^{(1)}+m_{30 \cdot w}^{(1)} = {F_{01 \cdot 0w}}/(P_{1\cdot 1w}-P_{1\cdot 0w}),\\
            & m_{01 \cdot w}^{(1)}+m_{11 \cdot w}^{(1)}+m_{00 \cdot w}^{(1)}+m_{10 \cdot w}^{(1)}= {F_{11 \cdot 1w}}/(P_{1\cdot 1w}-P_{1\cdot 0w}),\\ 
            & m_{21 \cdot w}^{(1)}+m_{31 \cdot w}^{(1)} +m_{20 \cdot w}^{(1)}+m_{30 \cdot w}^{(1)}=  {F_{01 \cdot 1w}}/ (P_{1\cdot 1w}-P_{1\cdot 0w}),\\  
            & m_{00 \cdot w}^{(1)}+m_{10 \cdot w}^{(1)}+m_{20 \cdot w}^{(1)}+m_{30 \cdot w}^{(1)}=1 ,\\
            & \sum_{j=0}^1\sum_{i=0}^3 m_{ij \cdot w}^{(1)}+m_{\cdot 3w}^{(1)} =  {1}/(P_{1\mid 1w}-P_{1\mid 0w}), \\       
             & m_{ij \cdot w}^{(1)} \geq 0\,\,(i=0,1,2,3,j=0,1), m_{\cdot 3w}^{(1)}\geq 0.
            \end{aligned}
\right.
\end{eqnarray}
Solving the above linear programming problem yields,
\begin{equation}
\label{b_LD_mon}
L_{1,m}(y,w)\leq m_{21\cdot w}^{(1)}-m_{11\cdot w}^{(1)}\leq R_{1,m}(y,w), 
\end{equation}
where
\begin{eqnarray*}
R_{1,m}(y,w)= \min \left\{
\frac {F_{01 \cdot 1w}}{P_{1\cdot 1w}-P_{1\cdot 0w} },
1
\right\}\,\mbox{and}\,L_{1,m}(y,w)= \max \left\{
\frac {-F_{11 \cdot 1w}}{P_{1\cdot 1w}-P_{1\cdot 0w} }, 
      -1
\right\}.
\end{eqnarray*}
Similarly, since  $\Delta_1 ={E\{(P_{1\cdot 1W}-P_{1\cdot 0W})\Delta_{1\cdot W})\}}/{E(P_{1\cdot 1W}-P_{1\cdot 0W})}$ we can obtain the following bounds
\begin{equation}
 \int_{-\infty}^{+\infty}  L_{1,m}(y,w)\, dy\leq \Delta_{1\cdot w} \leq \int_{-\infty}^{+\infty}  R_{1,m}(y,w)\, dy,
\end{equation} 
and
\begin{equation}
L_{1,m} \leq \Delta_{1} \leq R_{1,m},
\end{equation}
where
\begin{eqnarray*}
\begin{aligned}
&L_{1,m}=\frac{E\Big\{(P_{1\cdot 1W}-P_{1\cdot 0W})\int_{-\infty}^{+\infty}  L_{1,m}(y,W)\, dy\Big\}}{E(P_{1\cdot 1W}-P_{1\cdot 0W})},\\
&R_{1,m}=\frac{E\Big\{(P_{1\cdot 1W}-P_{1\cdot 0W})\int_{-\infty}^{+\infty}  R_{1,m}(y,W)\, dy\Big\}}{E(P_{1\cdot 1W}-P_{1\cdot 0W})}.
\end{aligned}
\end{eqnarray*} 
\end{proof}

\subsection{Proof of Theorem~\ref{thm-Deltas-sd}}
\label{Appendix: thm-Deltas-sd}
\begin{proof}
    Notably, Assumption~\ref{assu4} is equivalent to
\begin{eqnarray*}
    \left\{\begin{aligned}        
          & \pi_{1\cdot w}(F_{00\cdot w}+F_{10\cdot w})-
          \pi_{0\cdot w}(F_{01\cdot w}+F_{11\cdot w})\leq 0,\\
          &  \pi_{2\cdot w}(F_{00\cdot w}+F_{20\cdot w})-
          \pi_{0\cdot w}(F_{02\cdot w}+F_{22\cdot w})\leq 0.
    \end{aligned}
    \right.
\end{eqnarray*}
Recall the notation used in (\ref{linear programming 1}), (\ref{fracPf}), (\ref{fracPf1}) and (\ref{fracPf2}), and recall that $F_{00\cdot w}+F_{10\cdot w}+F_{01\cdot w}+F_{11\cdot w}=F_{01\cdot 0w}$, $F_{20\cdot w}+F_{20\cdot w}+F_{02\cdot w}+F_{22\cdot w}=F_{01\cdot 1w}$, $\pi_{0\cdot w}+\pi_{1\cdot w}=P_{1\cdot0w }$ and $\pi_{0\cdot w}+\pi_{2\cdot w}=P_{1\cdot1w }$; then, the above inequalities are equivalent to
\begin{eqnarray}\label{sd_add}
\scriptstyle{
    \left\{\begin{aligned}        
          & f_{00 \cdot w}^{(0)}+f_{10 \cdot w}^{(0)} \leq \frac{F_{11 \cdot 1w}}{P_{1\cdot 1w}},\\
          &  f_{00 \cdot w}^{(0)}+f_{20 \cdot w}^{(0)} \leq  \frac{F_{11 \cdot 0w}}{P_{1\cdot 0w}}.
    \end{aligned}
    \right.
    \,\,\mbox{or}\,\,
    \left\{\begin{aligned}        
          &f_{01 \cdot w}^{(1)}+f_{11 \cdot w}^{(1)} \geq \frac{F_{11 \cdot 1w}}{P_{1\cdot 1w}},\\
          &  f_{00 \cdot w}^{(1)}+f_{20 \cdot w}^{(1)} \leq  \frac{\pi_{0\cdot w} F_{11 \cdot 0w}}{\pi_{1\cdot w} P_{1\cdot 0w}}.
    \end{aligned}
    \right.
    \,\,\mbox{or}\,\,
    \left\{\begin{aligned}
        & f_{00 \cdot w}^{(2)}+f_{10 \cdot w}^{(2)} \leq \frac{\pi_{0\cdot w} F_{11 \cdot 1w}}{\pi_{2\cdot w} P_{1\cdot 1w}},\\
          & f_{02 \cdot w}^{(2)}+f_{22 \cdot w}^{(2)} \geq \frac{F_{11 \cdot 0w}}{P_{1\cdot 0w}}.
    \end{aligned}
    \right.
}
\end{eqnarray}
By adding the left constraint of (\ref{sd_add}) to the problem given in (\ref{fracPf}), under Assumptions 1, 2 and 4, we  obtain
\begin{equation}
    \int_{-\infty}^{\infty}L_{0,\mbox{sd}}(y,w\mid \pi_{2\cdot w}^{\scriptstyle\max})\,dy \leq \mbox{SACE}_w \leq \int_{-\infty}^{\infty}R_{0,\mbox{sd}}(y,w\mid \pi_{2\cdot w}^{\scriptstyle\max})\,dy,
\end{equation}
and
\begin{equation}\label{delda_0_sd}
  L_{0,\mbox{sd}}\leq \mbox{SACE}\leq  R_{0,\mbox{sd}},
\end{equation}
where
\begin{eqnarray*}
    \begin{aligned}
        & L_{0,\mbox{sd}}(y,w\mid \pi_{2\cdot w}^{\scriptstyle\max})= \max \left\{\frac {-F_{11 \cdot 1w}}{P_{1\cdot 1w}}, \frac {F_{01 \cdot 1w}}{P_{1\cdot 1w}} - \frac {F_{01 \cdot 0w}}{P_{1\cdot 0w}-\pi_{2\cdot w}^{\scriptstyle\max}}
\right\},\\
        & L_{0,\mbox{sd}}=\frac{E\left\{(P_{1\cdot 0W}-\pi_{2\cdot W}^{\scriptstyle\max}) \int_{-\infty}^{+\infty} L_{0,\mbox{sd}}(y,W\mid \pi_{2\cdot W}^{\scriptstyle\max})dy\right\}}{E(P_{1\cdot 0W}-\pi_{2\cdot W}^{\scriptstyle\max})}, \\
        & R_{0,\mbox{sd}}(y,w\mid \pi_{2\cdot w}^{\scriptstyle\max}) = \min \left\{
\frac {F_{11 \cdot 0w}}{P_{1\cdot 0w}}, \frac {-F_{01 \cdot 0w}}{P_{1\cdot 0w}} + \frac {F_{01 \cdot 1w}}{P_{1\cdot 0w}-\pi_{2\cdot w}^{\scriptstyle\max}} \right\},\\
        & R_{0,\mbox{sd}}=\frac{E\left\{ (P_{1\cdot 0W}-\pi_{2\cdot w}^{\scriptstyle\max})\int_{-\infty}^{+\infty}R_{0,\mbox{sd}}(y,W\mid \pi_{2\cdot W}^{\scriptstyle\max})dy\right\}}{E(P_{1\cdot 0W}-\pi_{2\cdot W}^{\scriptstyle\max})}.
    \end{aligned}
\end{eqnarray*}
Solving the new programming problem (\ref{fracPf1}) by adding the middle constraint of (\ref{sd_add}), we obtain
\begin{equation}
\int_{-\infty}^{\infty} L_{1,\mbox{sd}}(y,w\mid \pi_{2\cdot w}^{\scriptstyle\min})\,dy\leq \Delta_{1\cdot w}\leq \int_{-\infty}^{\infty}R_{1,\mbox{sd}}(y,w\mid \pi_{2\cdot w}^{\scriptstyle\min})\,dy, 
\end{equation}
and
\begin{equation}
\label{b_LD_stoc}
 L_{1,\mbox{sd}}\leq \Delta_{1}\leq R_{1,\mbox{sd}}, 
\end{equation}
where
\begin{eqnarray*}
\begin{aligned}
&L_{1,\mbox{sd}}(y,w|\pi_{2\cdot w}^{\scriptstyle\min})= \max \left\{
\frac {-F_{11 \cdot 1w}}{P_{1\cdot 1w}-P_{1\cdot 0w}+\pi_{2\cdot w}^{\scriptstyle\min}}, 
       -1
\right\},\,\,\,\,R_{1,\mbox{sd}}(y,w|\pi_{2\cdot w}^{\scriptstyle\min})= \frac {F_{01 \cdot 1w}}{P_{1\cdot 1w}},\\
& L_{1,\mbox{sd}} = \frac{E\{(P_{1\cdot 1W}-P_{1\cdot 0W}+\pi_{2\cdot W}^{\scriptstyle\min})\int_{-\infty}^\infty L_{1,\mbox{sd}}(y,W|\pi_{2\cdot W}^{\scriptstyle\min})\,dy\}}{E(P_{1\cdot 1W}-P_{1\cdot 0W}+\pi_{2\cdot W}^{\scriptstyle\min})} ,\\
& R_{1,\mbox{sd}} = \frac{E\{(P_{1\cdot 1W}-P_{1\cdot 0W}+\pi_{2\cdot W}^{\scriptstyle\min})\int_{-\infty}^\infty R_{1,\mbox{sd}}(y,W|\pi_{2\cdot W}^{\scriptstyle\min})\,dy\}}{E(P_{1\cdot 1W}-P_{1\cdot 0W}+\pi_{2\cdot W}^{\scriptstyle\min})}. 
\end{aligned}
\end{eqnarray*}
When considering the case in which the right constraint of (\ref{sd_add}) is added to the programming problem in
(\ref{fracPf2}), we obtain
\begin{equation}
\int_{-\infty}^{\infty} L_{2,\mbox{sd}}(y,w\mid \pi_{2\cdot w}^{\scriptstyle\min})\,dy\leq \Delta_{2\cdot w}\leq \int_{-\infty}^{\infty}R_{2,\mbox{sd}}(y,w\mid \pi_{2\cdot w}^{\scriptstyle\min})\,dy, 
\end{equation}
and
\begin{equation}
\label{b_DL_stoc}
 L_{2,\mbox{sd}}\leq \Delta_{2}\leq R_{2,\mbox{sd}}, 
\end{equation}
where
\begin{eqnarray*}
\begin{aligned}
&L_{2,\mbox{sd}}(y,w|\pi_{2\cdot w}^{\scriptstyle\min})= 
\frac {-F_{01 \cdot 0w}}{P_{1\cdot 0w}},
&R_{2,\mbox{sd}}(y,w|\pi_{2\cdot w}^{\scriptstyle\min})= \min \left\{
\frac {F_{11 \cdot 0w}}{\pi_{2\cdot w}^{\scriptstyle\min}}, 1
\right\},\\
& L_{2,\mbox{sd}} = \frac{E\{\pi_{2\cdot W}^{\scriptstyle\min}\int_{-\infty}^\infty L_{2,\mbox{sd}}(y,W)\,dy\}}{E(\pi_{2\cdot W}^{\scriptstyle\min})},&
R_{2,\mbox{sd}} = \frac{E\{\pi_{2\cdot W}^{\scriptstyle\min}\int_{-\infty}^\infty R_{2,\mbox{sd}}(y,W|\pi_{2\cdot W}^{\scriptstyle\min})\,dy\}}{E(\pi_{2\cdot W}^{\scriptstyle\min})}. 
\end{aligned}
\end{eqnarray*} 
\end{proof}

\subsection{Proof of Theorem~\ref{thm-Deltas-m-sd}}
\label{Appendix: thm-Deltas-m-sd}
\begin{proof}
   We first study the strengthening of the linear program for the SACE by adding the left constraint of (\ref{sd_add}) to (\ref{fracPf0_mon}). Then, the bounds on the SACE under Assumptions 1-4 can be obtained as follows:
\begin{equation}
    \int_{-\infty}^{\infty}L_{0,\mbox{m},\mbox{sd}}(y,w)\,dy \leq \mbox{SACE}_w \leq \int_{-\infty}^{\infty}R_{0,\mbox{m},\mbox{sd}}(y,w)\,dy,
\end{equation}
and
\begin{equation}
  L_{0,\mbox{m},\mbox{sd}}\leq \mbox{SACE}\leq  R_{0,\mbox{m},\mbox{sd}},
\end{equation}
where
\begin{eqnarray*}
    \begin{aligned}
        & L_{0,\mbox{m,sd}}(y,w)= \max \left\{\frac {F_{01 \cdot 1w}}{P_{1 \cdot 1w}}-\frac {F_{01 \cdot 0w}}{P_{1\cdot 0w}}, \frac {F_{11 \cdot 0w}-F_{11 \cdot 1w}}{P_{1\cdot 0w}}\right\},\\
        & L_{0,\mbox{m,sd}}=\frac{E\left\{P_{1\cdot 0W}\int_{-\infty}^{+\infty} L_{0,\mbox{m,sd}}(y,W)dy\right\}}{E(P_{1\cdot 0W})}, \\
        & R_{0,\mbox{m,sd}}(y,w) = \min\left\{\frac {F_{11 \cdot 0w}}{P_{1\cdot 0w}}, \frac {F_{01 \cdot1w}-F_{01 \cdot 0w}}{P_{1\cdot 0w}}\right\},\\
        & R_{0,\mbox{m,sd}}=\frac{E\left\{ P_{1\cdot 0W}\int_{-\infty}^{+\infty}R_{0,\mbox{m,sd}}(y,W)dy\right\}}{E(P_{1\cdot 0W})}.
    \end{aligned}
\end{eqnarray*}
Consider the linear programming problem in (\ref{fracPf1_mon}), which is enriched by additionally introducing the middle constraint of (\ref{sd_add}). Then, the bounds in the `protected' stratum are
\begin{equation}
    \int_{-\infty}^{\infty}L_{1,\mbox{m},\mbox{sd}}(y,w)\,dy \leq \Delta_{1\cdot w} \leq \int_{-\infty}^{\infty}R_{1,\mbox{m},\mbox{sd}}(y,w)\,dy,
\end{equation}
and
\begin{equation}
  L_{1,\mbox{m},\mbox{sd}}\leq \Delta_1\leq  R_{1,\mbox{m},\mbox{sd}},
\end{equation}
where
\begin{eqnarray*}
    \begin{aligned}
       & L_{1,\mbox{m,sd}}(y,w)= \max \left\{
\frac {-F_{11 \cdot 1w}}{P_{1\cdot 1w}-P_{1\cdot 0w}}, 
       -1
\right\}, \,\,R_{1,\mbox{m,sd}}(y,w) = 
\frac {F_{01 \cdot 1w}}{P_{1\cdot 1w}},\\
&L_{1,\mbox{m,sd}}=\frac{E\left\{(P_{1\cdot 1W}-P_{1\cdot 0W})\int_{-\infty}^\infty L_{1,\mbox{m,sd}}(y,W)\,dy\right\}}{E(P_{1\cdot 1W}-P_{1\cdot 0W})},\\
&R_{1,\mbox{m,sd}}=\frac{E\left\{(P_{1\cdot 1W}-P_{1\cdot 0W})\int_{-\infty}^\infty R_{1,\mbox{m,sd}}(y,W)\,dy\right\}}{E(P_{1\cdot 1W}-P_{1\cdot 0W})}.
    \end{aligned}
\end{eqnarray*} 
\end{proof}

\section{\label{Appendix: extension}Extension: ACE and CACE}

In fact, our proposed approach is not limited to the case where the outcome variable is truncated by death, the analogous approach of section~3 can be applied to bound ACE and complier average causal effects (CACE). Next we give the general process of extension. 

As summarized by \citet{2018Swanson}, the assigned treatment $Z$ can be viewed as instrument variables, $D$ denoted the treatment received, and $Y(z,d)$ for $z,d\in\{0,1\}$ denoted potential outcome which depends both the instrument and the treatment. The population can be divided into four subpopulations based on the joint values of the potential treatment status under both values of the instrument $\{D(0), D(1)\}$ \citep{1996Angrist}: $g_0=\{D(0) = 0, D(1) = 0\}$, never-takers, individuals who never take the treatment regardless of the value of the instrument; $g_1=\{D(0) = 0, D(1) = 1\}$, compliers, individuals who take the treatment only if they are exposed to the instrument (i.e., only if $Z = 1$); $g_2=\{D(0) = 1, D(1) = 0\}$, defiers, individuals who take the treatment only if they are not exposed to the instrument; $g_3=\{D(0) = 1, D(1) = 1\}$, always-takers, individuals who always take the treatment regardless of the value of the instrument. These subpopulations are called “principal strata” in \citet{2002Frangakis}. In order to obtain the Balke-Pearl bounds on ACEs, the exclusion restriction and ignorability assumptions are commonly required, that is, $Y(z,d)=Y(z',d)=Y(d)$ for all $z,z',d$, and $Z \perp \{Y(1),Y(0),D(1),D(0)\} \mid W, G, \mbox{\,\,with\,\,}G\in\{g_0,g_1,g_2,g_3\}$. Then the observed outcome $Y$ and the observed treatment $D$ satisfy that $Y=DY(1)+(1-D)Y(0)$ and $D=ZD(1)+(1-Z)D(0),$ respectively. Without causing confusion, the joint conditional probability distribution of $R$ and $G$ conditional on $W$ is still denoted in this section by $F_{ij \cdot w}\equiv \P(R=r_i, G=g_j\mid W=w),\,\,\, i,j=0, 1, 2, 3.$ Let $F_{ad\cdot zw}$ denote the observed conditional joint probability distribution of $Y$ and $D=d$ given $Z=z$ and $W=w$, that is, for $a,d,z\in\{0,1\},$
$F_{ad\cdot zw}= \P(I(Y\leq y)=a,D=d \mid Z=z,W=w).$
Similar to (\ref{SACEw}), both $\mbox{ACE}_w \equiv E\{Y(1)-Y(0)|W=w\}$ and $\mbox{CACE}_w \equiv E\{Y(1)-Y(0)|D(1)=1,D(0)=0, W=w\}$
can be rewritten as
\begin{equation}
\mbox{ACE}_w =\int_{-\infty}^{\infty}\left\{ (F_{20\cdot w}+F_{21\cdot w}+F_{22\cdot w}+F_{23\cdot w}\right) - \left(F_{10\cdot w}+F_{11\cdot w}+F_{12\cdot w}+F_{13\cdot w} )\right\} dy,
\end{equation} 
and
\begin{equation}
\mbox{CACE}_w =\int_{-\infty}^{\infty} (F_{21\cdot w}-F_{11\cdot w})/(F_{01\cdot w}+F_{11\cdot w}+F_{21\cdot w}+F_{31\cdot w}) dy.
\end{equation} 
In terms of the above exclusion restriction and ignorability assumptions, we can construct the following constraints:
\begin{eqnarray}
\label{linear programming ACEs}
\left\{
		\begin{aligned}
            &F_{11 \cdot 1w}=F_{01 \cdot w}+F_{03 \cdot w}+F_{11 \cdot w}+F_{13 \cdot w},\,\,\,\,
            F_{01 \cdot 1w}=F_{21 \cdot w}+F_{23 \cdot w}+F_{31 \cdot w}+F_{33 \cdot w},\\
            &F_{11 \cdot 0w}=F_{02 \cdot w}+F_{03 \cdot w}+F_{12\cdot w}+F_{13\cdot w},\,\,\,\,
            F_{01 \cdot 0w}=F_{22 \cdot w}+F_{23 \cdot w}+F_{3 2\cdot w}+F_{33\cdot w},\\
            &F_{10 \cdot 1w}=F_{00 \cdot w}+F_{02 \cdot w}+F_{20 \cdot w}+F_{22 \cdot w},\,\,\,\,
            F_{00 \cdot 1w}=F_{10 \cdot w}+F_{12 \cdot w}+F_{30 \cdot w}+F_{32 \cdot w},\\
            &F_{10 \cdot 0w}=F_{00 \cdot w}+F_{01 \cdot w}+F_{20\cdot w}+F_{21\cdot w},\,\,\,\,
            F_{00 \cdot 0w}=F_{10 \cdot w}+F_{11 \cdot w}+F_{3 0\cdot w}+F_{31\cdot w},\\
            &1=\sum_{i,j\in \{0,1,2,3\}} F_{ij \cdot w}, \,\,\,\,F_{ij \cdot w} \geq 0. \\
        \end{aligned}
\right.                
\end{eqnarray}

The sharp bounds on the integrand of $\mbox{ACE}_w$ can be obtained by solving a linear programming problem as follows:
\begin{equation*}
\min\,\mbox{or}\, \max\, \{\left (F_{20\cdot w}+F_{21\cdot w}+F_{22\cdot w}+F_{23\cdot w}\right) - \left(F_{10\cdot w}+F_{11\cdot w}+F_{12\cdot w}+F_{13\cdot w} \right)\},
\end{equation*}
subject to (\ref{linear programming ACEs}), and denoted by 
$\big [L_{\mbox{\scriptsize {ACE}}}(y,w),R_{\mbox{\scriptsize {ACE}}}(y,w)\big ]$, where 
\begin{eqnarray*}
L_{\mbox{\scriptsize {ACE}}}(y,w)=\max\left\{
    \begin{aligned}   
&F_{01\cdot1w}+F_{10\cdot0w}-1, \\
&2F_{01\cdot1w}+F_{11\cdot0w}+F_{10\cdot1w}+F_{10\cdot0w}-2,\\
&F_{01\cdot0w}+2F_{10\cdot0w}-F_{11\cdot1w}-F_{10\cdot1w}-1,\\
&F_{01\cdot1w}+F_{10\cdot1w}-1, \\
&F_{01\cdot0w}+F_{10\cdot0w}-1, \\
&F_{01\cdot1w}+2F_{10\cdot1w}-F_{11\cdot0w}-F_{10\cdot0w}-1, \\
&F_{11\cdot1w}+2F_{01\cdot0w}+F_{10\cdot1w}+F_{10\cdot0w}-2,\\
&F_{01\cdot0w}+F_{10.1w}-1
\end{aligned}
 \right\},
\end{eqnarray*}
and
\begin{eqnarray*}
R_{\mbox{\scriptsize {ACE}}}(y,w)=\min\left\{
    \begin{aligned} 
&F_{11\cdot1w}+F_{01\cdot1w}+F_{10\cdot1w}-F_{11\cdot0w},\\
&F_{11\cdot1w}+2F_{01\cdot1w}+2F_{10\cdot1w}-F_{11\cdot0w}-F_{10\cdot0w},\\
&F_{11\cdot1w}+F_{01\cdot0w}+F_{10\cdot1w}+F_{10\cdot0w}-F_{11\cdot0w},\\ 
&F_{01\cdot1w}+F_{10\cdot1w},\\
&F_{01\cdot0w}+F_{10\cdot0w},\\
&F_{01\cdot1w}+F_{11\cdot0w}+F_{10\cdot1w}+F_{10\cdot0w}-F_{11\cdot1w}, \\
&F_{11\cdot0w}+2F_{01\cdot0w}+2F_{10\cdot0w}-F_{11\cdot1w}-F_{10\cdot1w}, \\
&F_{11\cdot0w}+F_{01\cdot0w}+F_{10\cdot0w}-F_{11\cdot1w} 
\end{aligned}
\right\}.
\end{eqnarray*}
As a result, the sharp bounds on ACE are
\begin{equation}\label{bounds_ACE}
    E\left\{\int_{-\infty}^{\infty}L_{\mbox{\scriptsize {ACE}}}(y,W)\,dy\right\} \leq \mbox{ACE} \leq E\left\{\int_{-\infty}^{\infty}R_{\mbox{\scriptsize {ACE}}}(y,W)\,dy\right\}.
\end{equation}
When $Y$ is binary, (\ref{bounds_ACE}) can reduce to the bounds of \cite{1997Balke}. It is worth mentioning that (\ref{bounds_ACE}) allows outcome $Y$ and covariate $W$ to be continuous or discrete.
 
On the other hand, CACE which is also commonly known as local average treatment effects (LATE), has been extensively studied in the literature such as \citet{2010Richardson} and \citet{2017Huber}.
Intuitively, we can obtain the sharp bounds on the integrands of $CACE_w$ by solving the following optimization problem subject to (\ref{linear programming ACEs}):
\begin{equation*}
\min\,\mbox{or}\, \max \frac{F_{21\cdot w}-F_{11\cdot w}}{F_{01\cdot w}+F_{11\cdot w}+F_{21\cdot w}+F_{31\cdot w}}.
\end{equation*}
Here the denominator can be rewritten as $P_{1\cdot1w}-P_{1\cdot0w}+\pi_{2\cdot w}$ where $P_{1\cdot1w}=\P (D=1|Z=1,W=w), P_{1\cdot0w}=\P (D=1|Z=0,W=w)$ and $\pi_{2\cdot w}=\P(G=g_2|W=w)$ which is not identified. Note that $\pi_{2\cdot w}^{\scriptstyle\min}=max\{0,P_{1\cdot0w}-P_{1\cdot1w}\}.$ Let $\mathcal{W}_w=P_{1\cdot1w}-P_{1.0w}+\pi_{2\cdot w}^{\scriptstyle\min}.$ Following the proposed procedure in section~3, we have 
\begin{eqnarray}
        \begin{aligned}
            \frac{E\left\{\mathcal{W}_W\int_{-\infty}^{\infty}L_{\mbox{\scriptsize {CACE}}}(y,W\mid \pi_{2\cdot W}^{\scriptstyle\min})\,dy\right\}}{E(\mathcal{W}_W)}
     \leq \mbox{CACE} \leq\frac{E\left\{\mathcal{W}_W\int_{-\infty}^{\infty}R_{\mbox{\scriptsize {CACE}}}(y,W\mid \pi_{2\cdot W}^{\scriptstyle\min})\,dy\right\}}{E(\mathcal{W}_W)},  
        \end{aligned}           
\end{eqnarray}
where
\begin{eqnarray*}
L_{\mbox{\scriptsize {CACE}}}(y,w\mid \pi_{2\cdot w}^{\scriptstyle\min}) = \max\left\{
    \begin{aligned}  
&(F_{10\cdot0w}-F_{11\cdot1w}-F_{10\cdot1w})/\mathcal{W}_w, \\
&(F_{01\cdot1w}+F_{11\cdot0w}+F_{10\cdot0w}-1)/\mathcal{W}_w, \\ 
&(F_{11\cdot0w}+F_{01\cdot0w}+F_{10\cdot0w}-1)/\mathcal{W}_w, \\
&-F_{11\cdot1w}/\mathcal{W}_w
    \end{aligned}
\right\},
\end{eqnarray*}
and 
\begin{eqnarray*}
R_{\mbox{\scriptsize {CACE}}}(y,w\mid \pi_{2\cdot w}^{\scriptstyle\min}) = \min\left\{
    \begin{aligned} 
&F_{10\cdot0w}/\mathcal{W}_w, \\
&F_{01\cdot1w}/\mathcal{W}_w, \\
&(F_{11\cdot0w}+F_{01\cdot0w}+F_{10\cdot0w}-F_{11\cdot1w}-F_{10\cdot1w})/\mathcal{W}_w,\\
&(F_{11\cdot0w}+F_{10\cdot0w}-F_{11\cdot1w})/\mathcal{W}_w
    \end{aligned}
\right\}.
\end{eqnarray*}

\section{More simulation studies}
\label{Appendix: simu}
In the main text, we only present how sensitive the proposed method is to deviations from both Assumptions 3 and 4 and the case in which all assumptions hold. Violations of one of the assumptions are reported separately here.  Again, we considered both discrete and continuous responses $Y$.
\begin{figure}
  \centering
  \includegraphics[width=0.8\textwidth]{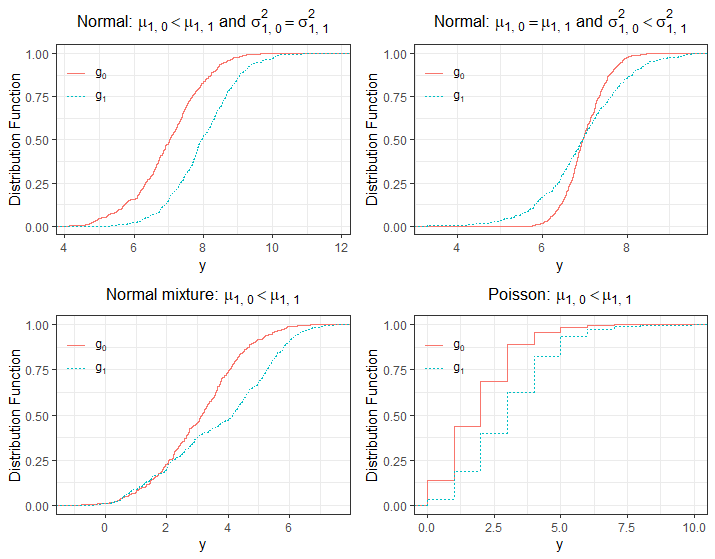}
  \caption{Distribution functions conditional on $G$  where Assumption 4 is violated. }
  \label{Vio_A4}
\end{figure}

\subsection{The case with Assumption 3 (monotonicity)}

Similar to Subsection \ref{no34}, we consider four cases of violation of Assumption~4, as shown in Figure \ref{Vio_A4} where $\mu_{1,0}<\mu_{1,1}$ or $\sigma_{1,0}^2<\sigma_{1,1}^2$ under the assumption that $\mu_{1,0}=\mu_{1,1}$. Note that the `harmed' stratum no longer exists under Assumption 3; that is, we can focus only on the case in which the distribution of $Y$ conditional on $Z=1,G=g_1$ does not stochastically dominate that conditional on $Z=1,G=g_0$. We set $\pi_{0} = 0.5$, $\pi_{1} = 0.3$ and $\pi_{3} = 0.2$. The true value of the SACE in the first two cases is $4$, that in the third case is $-0.5$ and that in the fourth case is $-5$. The simulated data are generated as follows:
\begin{itemize}
    \item {\bf The case of a normal $Y$ with $\mu_{1,0}<{\mu}_{1,1}$.} The parameters are $\mu_{1,0}=7,\mu_{1,1}=8,\mu_{0,0}=3$ and $\sigma_{1,0}^2=\sigma_{1,1}^2=\sigma_{0,0}^2=1.$
    \item {\bf The case of a normal $Y$ with $\sigma_{1,0}^2<\sigma_{1,1}^2$.} The parameters are $\sigma_{1,0}^2=1/2,\sigma_{1,1}^2=1,\sigma_{0,0}^2=1,$ $\mu_{1,0}=\mu_{1,1}=7,$ and $\mu_{0,0}=3.$  
    \item {\bf The case of a binary $Y$ with $\mu_{1,0}<{\mu}_{1,1}$.} The parameters are $\mu_{1,0}=0.2,\mu_{1,1}=0.3$ and $\mu_{0,0}=0.7.$
    \item {\bf The case of a Poisson $Y$ with $\mu_{1,0}<{\mu}_{1,1}$.} The parameters are $\mu_{1,0}=2,\mu_{1,1}=3$ and $\mu_{0,0}=7.$
\end{itemize}

Table \ref{table1-3} shows that when Assumption~4 is violated, our bounds are valid, and the sign of the SACE is accurately identified by the proposed bounds in the above four cases among 1000 replications.
\begin{table}[htb]
\renewcommand{\arraystretch}{0.8}
 \setlength{\belowcaptionskip}{0.2cm}     
 \centering\caption{The bounds on the SACE without Assumption~4.}
 \label{table1-3}
 \begin{tabular}{l|ccc}
           \hline
           \quad  & True value  &  Bounds  &  Coverage \\
           \hline
Normal: $\mu_{1,0}<\mu_{1,1},\sigma_{1,0}^2=\sigma_{1,1}^2$   & 4  &  $[3.868, 5.225]$ & 100\% \\          
Normal: $\mu_{1,0}=\mu_{1,1},\sigma_{1,0}^2<\sigma_{1,1}^2$ & 4 & $[3.530, 4.471]$ & 100\% \\    
Binary: $\mu_{1,0}<\mu_{1,1}$  & -0.5 & $[-0.700, -0.320]$ & 100\% \\
Poisson: $\mu_{1,0}<\mu_{1,1}$   & -5  &  $[-5.610, -3.695]$ & 100\% \\
           \hline    
\end{tabular}
\end{table}

\subsection{The case with Assumption 4 (stochastic dominance)}

Now, we assess the performance of the proposed method described in Subsection \ref{section:without M} when Assumption~3 (monotonicity) is violated but Assumption~4 (stochastic dominance) holds. We calculate numerous probabilities that event $\{G=g_k\}$ occurs, similarly Subsection \ref{no34}: $\pi_{1} = 0.3$, $\pi_{3} = 0.1$ and $\pi_{0} = 0.6 -\pi_{2}$ where $\pi_{2}$ is varied from $0.05$ to $0.25$. The true values of the SACE in the first two cases are $5$ and $4$, respectively; that in the third case is $0.7$, and that in the fourth case is $5$. 

\begin{figure}[htbp]
  \centering
  \includegraphics[width=0.8\textwidth]{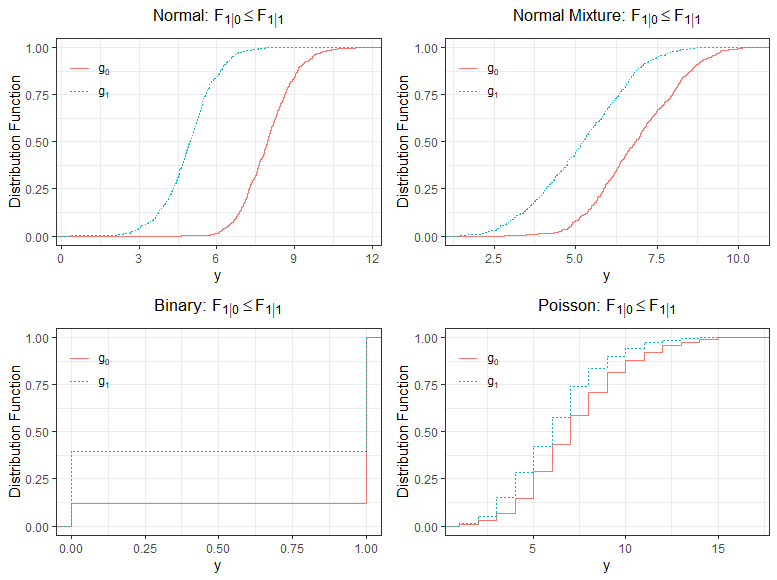}
  \caption{Distribution functions conditional on $G$  where Assumption 4 holds. }
  \label{Hold_A4}
\end{figure}

The simulated data are generated as follows:
\begin{itemize}
    \item {\bf The case of a normal $Y$ with $F_{1\mid 0}\leq F_{1|1}$ and $F_{0|0}\leq F_{0|2}$.} The parameters are $\mu_{1,0}=8,\mu_{1,1}=5,\mu_{0,0}=3,\mu_{0,2}=1$ with $\sigma_{1,0}^2=\sigma_{1,1}^2=\sigma_{0,0}^2=\sigma_{0,2}^2=1.$
    \item {\bf The case of a normal mixture $Y$ with $F_{1|0}\leq F_{1|1}$ and $F_{0|0}\leq F_{0|2}$.} 
    The parameters are
    $\omega_{1,0}=0.5,\omega_{1,1}=0.4,\mu_{1,0}^{(1)}=6,\mu_{1,0}^{(2)}=8,\mu_{1,1}^{(1)}=4,\mu_{1,1}^{(2)}=6$ with $\sigma^{2(1)}_{1,0}=\sigma^{2(2)}_{1,0}=\sigma^{2(1)}_{1,1}=\sigma^{2(2)}_{1,1}=1$ and $\omega_{0,0}=0.5,\omega_{0,2}=0.5,\mu_{0,0}^{(1)}=4,\mu_{0,0}^{(2)}=2,\mu_{0,2}^{(1)}=2,\mu_{0,2}^{(2)}=1$ with $\sigma^{2(1)}_{0,0}=\sigma^{2(2)}_{0,0}=\sigma^{2(1)}_{0,2}=\sigma^{2(2)}_{0,2}=1.$
    \item {\bf The case of a binary $Y$ with $F_{1|0}\leq F_{1|1}$ and $F_{0|0}\leq F_{0|2}$.} The parameters are $\mu_{1,0}=0.9,\mu_{1,1}=0.6, \mu_{0,0}=0.2$ and $\mu_{0,2}=0.1.$
    \item {\bf The case of a Poisson $Y$ with $F_{1|0}\leq F_{1|1}$ and $F_{0|0}\leq F_{0|2}$.} The parameters are $\mu_{1,0}=7,\mu_{1,1}=6,\mu_{0,0}=2$ and $\mu_{0,2}=1.$    
\end{itemize}
To illustrate Assumption~4, the distribution functions given $G$ in various cases are shown in Figure \ref{Hold_A4} where the distributions given $G=g_1$ dominate those given $G=g_0.$

\begin{figure}[htbp]
  \centering
  \includegraphics[width=0.8\textwidth]{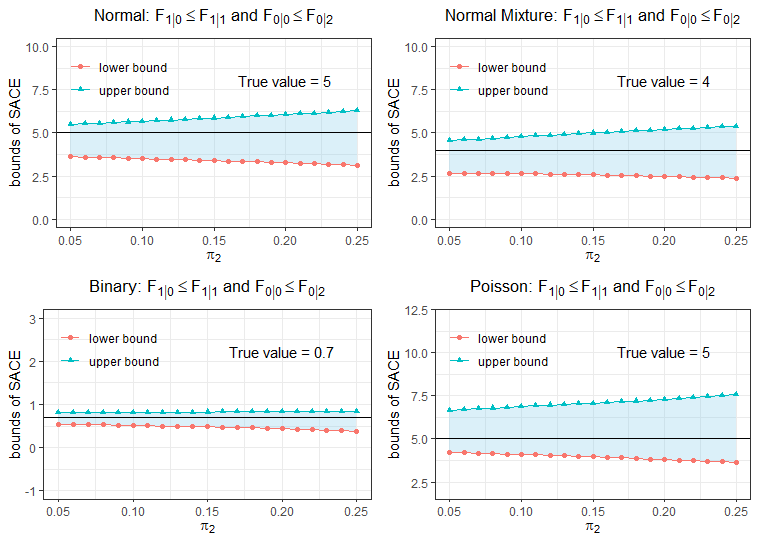}
  \caption{Bounds on the SACE with Assumption~4.}\label{fig:simu_124}
\end{figure}

Figure \ref{fig:simu_124} shows the averages of the estimated lower and upper bounds over 1000 replications when $\pi_2$ is varied from 0.05 to 0.25. The bounds we propose do not cover zero and thus can correctly identify the sign of the SACE. By examining this process of solving for the upper and lower bounds on the SACE, we find that the upper/lower bounds are achieved at different vertices for the normal, normal mixture and Poisson cases; specifically, the upper and lower bounds are achieved at vertices $\left\{F_{11.0}/P_{1.0}\right\}$ and vertex $\left\{F_{01.1}/P_{1.1}-F_{01.0}/(P_{1.0}-\pi_2)\right\}$, respectively.

\section{Further analysis of real data }
\label{Appendix: real data}

Based on the description of the NSW dataset, we specify that $\mbox{`Black'}=1$ if the individual was Black, $\mbox{`Married'}=1$ if the individual was married, $\mbox{`Unem75'}=1$ if the individual was unemployed for all of 1975, $\mbox{`Hisp'}=1$ if the individual was Hispanics, $\mbox{`Nodegree'}=1$ if the individual didn't have a high school degree. The bounds given for each covariate are shown in Tables \ref{table: con_cov_g0}-\ref{table: bin_cov_g1}. Notably, under Assumptions \ref{assu1}-\ref{assu4}, regardless of whether the considered covariate is discrete or continuous, the lower bound on the SACE is always greater than zero, which implies a positive effect. Specifically, both the married subgroup and the unmarried subgroup show a positive effect, but our results indicate that job training has a greater wage effect on the married subgroup than on the unmarried subgroup; family economic pressure may be one important reason for this difference. Consistent with common sense, a non-Hispanics individual ($\mbox{`Hisp'}=0$) or individuals with work experience ($\mbox{`Unem75'}=0$) or a high school degree ($\mbox{`Nodegree'}=0$) are more likely to find new jobs and earn higher wages. More precisely, for the covariates `Black', the wage effect of job training is positive for a Black individual ($\mbox{`Black'}=1$) as Black workers may have been more eager to find new jobs to make ends meet; otherwise, it is not clear whether the effect is positive or negative. 
\begin{table}[htb]
\renewcommand{\arraystretch}{0.8}
 \setlength{\belowcaptionskip}{0.2cm}     
 \centering\caption{Bounds on the SACE on earnings (measured on a natural log scale) for the NSW data, considering the continuous covariates `Education' and `Age'.}
\label{table: con_cov_g0}
\begin{tabular}{lccc}
          \hline
          Assumption & Without covariate & Education & Age \\
          \hline
          None & $[-1.289,1.410]$ & $[-1.333, 1.390]$ & $[-1.291, 1.426]$\\
          Monotonicity & $[-0.167,0.406]$ & $[-0.202, 0.374]$ & $[-0.151, 0.397]$  \\
          Stochastic Dominance & $[-0.516, 0.875]$ & $[-0.548, 0.852]$ & $[-0.514, 0.884]$  \\
          Both & 
          $[0.080,0.406]$ & $[0.049, 0.374]$ & $[0.088, 0.397]$  \\
          \hline
\end{tabular}
\end{table}

\begin{table}
\label{SACE_each}
\renewcommand{\arraystretch}{0.8}
 \setlength{\belowcaptionskip}{0.2cm}     
 \centering\caption{Bounds on the SACE on earnings (measured on a natural log scale) for the NSW data, given the covariates `Black', `Unem75', `Married', `Hisp' and `Nodegree'.}
 \scalebox{0.9}{
 \begin{tabular}{lcccc} 
          \hline
          Assumption & Without cov. & Black=0 & Black=1 & Overall range with cov.\\
          \hline
          None & $[-1.289,1.410]$ & $[-0.647, 0.329]$ & $[-1.489, 1.760]$ & $[-1.228, 1.316]$\\
          Mon. & $[-0.167,0.406]$ & $[-0.383, 0.088]$ & $[-0.101, 0.514]$ & $[-0.161, 0.424]$ \\
          Stoc. domin. & $[-0.516, 0.875]$ & $[-0.335, 0.223]$ & $[-0.569, 1.055]$ & $[-0.496, 0.797]$ \\
          Both & $[0.080,0.406]$ & $[-0.182, 0.088]$ & $[0.165, 0.514]$ & $[0.091, 0.424]$ \\
          \hline
          \quad & Without cov. & Unem75=0 & Unem75=1 & Overall range with cov.\\
          \hline
          None & $[-1.289,1.410]$ & $[-1.102, 1.219]$ & $[-1.419, 1.544]$ & $[-1.288, 1.410]$ \\
          Mon. & $[-0.167,0.406]$ & $[-0.229, 0.480]$ & $[-0.114, 0.339]$ & $[-0.155, 0.390]$ \\
          Stoc. domin. & $[-0.516, 0.875]$ & $[-0.415, 0.814]$ & $[-0.538, 0.916]$ & $[-0.514, 0.874]$ \\
          Both & $[0.080,0.406]$ & $[0.080, 0.480]$ & $[0.080, 0.339]$ & $[0.080, 0.390]$ \\
          \hline 
          \quad & Without cov. & Married=0 & Married=1 & Overall range with cov.\\
          \hline
          None & $[-1.289,1.410]$ & $[-1.331, 1.462]$ & $[-1.129, 1.211]$ & $[-1.292,1.272]$\\
          Mon. & $[-0.167,0.406]$ & $[-0.108, 0.331]$ & $[-0.401, 0.651]$ & $[-0.154, 0.382]$ \\
          Stoc. domin. & $[-0.516, 0.875]$ & $[-0.551, 0.880]$ & $[-0.357, 0.874]$ & $[-0.514, 0.879]$ \\
          Both & $[0.080,0.406]$ & $[0.080, 0.331]$ & $[0.080, 0.651]$ & $[0.080, 0.382]$ \\
          \hline
          \quad & Without cov. & Hisp=0 & Hisp=1 & Overall range with cov.\\
          \hline          
          None & $[-1.289,1.410]$ & $[-1.397, 1.564]$ & $[-0.293, 0.386]$ & $[-1.202, 1.355]$ \\
          Mon. & $[-0.167,0.406]$ & $[-0.153, 0.456]$ & $[-0.293, 0.386]$ & $[-0.169, 0.448]$ \\
          Stoc. domin. & $[-0.516, 0.875]$ & $[-0.548, 0.955]$ & $[0.002, 0.386]$ & $[-0.451, 0.854]$ \\
          Both & $[0.080,0.406]$ & $[0.110, 0.456]$ & $[0.002, 0.386]$ & $[0.098, 0.448]$ \\
          \hline
          \quad & Without cov. & Nodegree=0 & Nodegree=1 & Overall range with cov.\\
          \hline   
          None &$[-1.289,1.410]$ & $[-1.057, 1.417]$ & $[-1.388, 1.376]$ & $[-1.307,1.386]$\\
          Mon. & $[-0.167,0.406]$ & $[-0.104,0.594]$ & $[-0.204,0.302]$ & $[-0.182,0.366]$\\
          Stoc. domin. & $[-0.516, 0.875]$ &	$[-0.324, 0.977]$ &	$[-0.613, 0.802]$ & $[-0.542, 0.845]$\\
          Both & $[0.080,0.406]$ & $[0.197 ,0.594]$ & $[0.013,0.302]$ &	$[0.053,0.366]$\\
          \hline          
\end{tabular}}
\end{table}


\begin{table}
\renewcommand{\arraystretch}{0.8}
 \setlength{\belowcaptionskip}{0.2cm}     
 \centering\caption{Bounds on the ACE on earnings (measured on a natural log scale) within the `protected' stratum for the NSW data, considering the binary covariates `Black', `Unem75', `Married' , `Hisp' and `Nodegree'.}
  \label{table: bin_cov_g1}
\scalebox{0.9}{
 \begin{tabular}{lcccc}
          \hline
          Assumption & Without cov. & Black=0 & Black=1 & Overall range with cov. \\
          \hline
          None & $[-4.262, 4.773]$ & $[-4.157, 4.847]$ & $[-4.328, 4.742]$ & $[-4.301, 4.759]$ \\
          Mon. & $[-4.262, 4.773]$ & $[-4.157, 4.847]$ & $[-4.328, 4.742]$ & $[-4.301, 4.759]$ \\
          Stoc. domin. & $[-4.262, 3.332]$ & $[-4.157, 3.329]$ & $[-4.328, 3.332]$ & $[-4.301, 3.332]$ \\
          Both & $[-4.262, 3.332]$ & $[-4.157, 3.329]$ & $[-4.328, 3.332]$ & $[-4.301, 3.332]$  \\
          \hline
          \quad & Without cov. & Unem75=0 & Unem75=1 & Overall range with cov. \\
          \hline
          None & $[-4.262, 4.773]$ & $[-4.088, 4.668]$ & $[-4.425, 4.881]$ & $[-4.254, 4.773]$ \\
          Mon. & $[-4.262, 4.773]$ & $[-4.088, 4.668]$ & $[-4.425, 4.881]$ & $[-4.254, 4.773]$ \\
          Stoc. domin. & $[-4.262, 3.332]$  & $[-4.088, 3.332]$ & $[-4.425, 3.332]$ & $[-4.254, 3.332]$ \\
          Both & $[-4.262, 3.332]$  & $[-4.088, 3.332]$ & $[-4.425, 3.332]$ & $[-4.254, 3.332]$  \\
          \hline
          \quad & Without cov. & Married=0 & Married=1 & Overall range with cov. \\
          \hline
          None & $[-4.262, 4.773]$ & $[-4.444, 4.895]$ & $[-3.691, 4.455]$ & $[-4.144, 4.719]$ \\
          Mon. &$[-4.262, 4.773]$ & $[-4.444, 4.895]$ & $[-3.691, 4.455]$ & $[-4.144, 4.719]$ \\
          Stoc. domin. & $[-4.262, 3.332]$ & $[-4.444, 3.332]$ & $[-3.691, 3.332]$ &  $[-4.144, 3.332]$ \\
          Both & $[-4.262, 3.332]$ & $[-4.444, 3.332]$ & $[-3.691, 3.332]$ &  $[-4.144, 3.332]$  \\
          \hline
          \quad & Without cov. &  Hisp=0 & Hisp=1 & Overall range with cov. \\
          \hline
          None & $[-4.262, 4.773]$ & $[-4.215, 4.744]$ & $[-4.125, 4.690]$ & $[-4.204, 4.737]$ \\
          Mon. & $[-4.262, 4.773]$ & $[-4.215, 4.744]$ & $[-4.125, 4.690]$ & $[-4.204, 4.737]$ \\
          Stoc. domin. & $[-4.262, 3.332]$  & $[-4.215, 3.332]$ & $[-4.125, 3.332]$ & $[-4.204, 3.332]$ \\
          Both & $[-4.262, 3.332]$  & $[-4.215, 3.332]$ & $[-4.125, 3.332]$ & $[-4.204, 3.332]$ \\
          \hline
          \quad & Without cov. &  Nodegree=0 & Nodegree=1 & Overall range with cov. \\
          \hline
          None & $[-4.262, 4.773]$ & $[-4.017, 4.561]$ & $[-4.382, 4.734]$ & $[-4.273, 4.682]$ \\
          Mon. & $[-4.262, 4.773]$ & $[-4.017, 4.561]$ & $[-4.382, 4.734]$ & $[-4.273, 4.682]$ \\
          Stoc. domin. & $[-4.262, 3.332]$  & $[-4.017, 3.210]$ & $[-4.382, 3.264]$ & $[-4.273, 3.248]$ \\
          Both & $[-4.262, 3.332]$  & $[-4.017, 3.210]$ & $[-4.382, 3.264]$ & $[-4.273, 3.248]$ \\
          \hline
\end{tabular}}
\end{table}


\end{appendices}
\end{sloppypar}
\end{document}